\pdfoutput=1
\documentclass[
  reprint,
  superscriptaddress,
 amsmath,amssymb,
 aps,
prb,
]{revtex4-1}

\usepackage{graphicx}
\usepackage{dcolumn}
\usepackage{bm}
\usepackage{color}

\bmdefine{\bdi}{i}
\bmdefine{\bdj}{j}
\bmdefine{\bdx}{x}
\bmdefine{\bdy}{y}
\bmdefine{\bdr}{r}
\bmdefine{\bdR}{R}
\bmdefine{\bdS}{S}
\bmdefine{\bdL}{L}
\bmdefine{\bdJ}{J}
\bmdefine{\bdA}{A}
\bmdefine{\bdE}{E}
\bmdefine{\bdD}{D}
\bmdefine{\bdQ}{Q}
\bmdefine{\bdq}{q}
\bmdefine{\bdzero}{0}
\bmdefine{\bdv}{v}
\bmdefine{\bde}{e}
\bmdefine{\bddelta}{\delta}

\begin{document}

\title{
  Magnon drag induced by magnon-magnon interactions characteristic of noncollinear magnets
}

\author{Naoya Arakawa}
\email{arakawa@phys.chuo-u.ac.jp}
\affiliation{The Institute of Science and Engineering,
  Chuo University, Bunkyo, Tokyo, 112-8551, Japan}


\begin{abstract}
  A noncollinear magnet consists of
  the magnetic moments forming 
  a noncollinear spin structure. 
  Because of this structure,
  the Hamiltonian of magnons acquires the cubic terms.
  Although the cubic terms are the magnon-magnon interactions
  characteristic of noncollinear magnets,
  their effects on magnon transport have not been clarified yet.
  Here we show that in a canted antiferromagnet 
  the cubic terms cause a magnon drag that
  magnons drag magnon spin current and heat current,
  which can be used to enhance these currents
  by tuning a magnetic field.
  For a strong magnetic field, 
  we find that
  the cubic terms
  induce low-temperature peaks of a spin-Seebeck coefficient,
  a magnon conductivity, and a magnon thermal conductivity,
  and that each value is one order of magnitude larger 
  than the noninteracting value.
  This enhancement is mainly due to 
  the magnetic field dependence of the coupling constant of the cubic terms
  through the magnetic-field dependent canting angle.
  Our magnon drag offers a way for controlling the magnon currents
  of noncollinear magnets via the many-body effect. 
\end{abstract}
\maketitle


\section{Introdution}

Drag effects are nonequilibrium many-body effects.
In contrast to electronic and magnetic properties,
transport properties are essentially nonequilibrium
because a current makes a system out of equilibrium.
Then,
transport properties are often described by a theory without interactions,
but they are drastically changed by the effects of interactions,
many-body effects.
One of such examples is the phonon drag~\cite{PD-exp,PD-theory}.
The total momentum of electrons or phonons is not conserved
with the electron-phonon interaction.
As a result,
phonons drag an electron charge current
in the Seebeck effect~\cite{PD-theory,PD-theory2,Ogata}.
This phonon drag sometimes causes a peak
of the Seebeck coefficient~\cite{PD-exp,PD-theory2,PD-exp2}.
The other drag effects,
including the Coulomb drag~\cite{CD-exp,CD-theory},
the spin-Coulomb drag~\cite{SCD-theory,SCD-exp,NA-SCD},
the spin drag~\cite{SD1,SD2,SD3},
and the standard magnon drag~\cite{MD-theory,MD-exp1,MD-exp2,MD-exp3},
can be similarly understood.
Since the drag effects change transport properties qualitatively,
understanding their effects is
one of the central issues in condensed-matter physics. 

\begin{figure}
  \includegraphics[width=80mm]{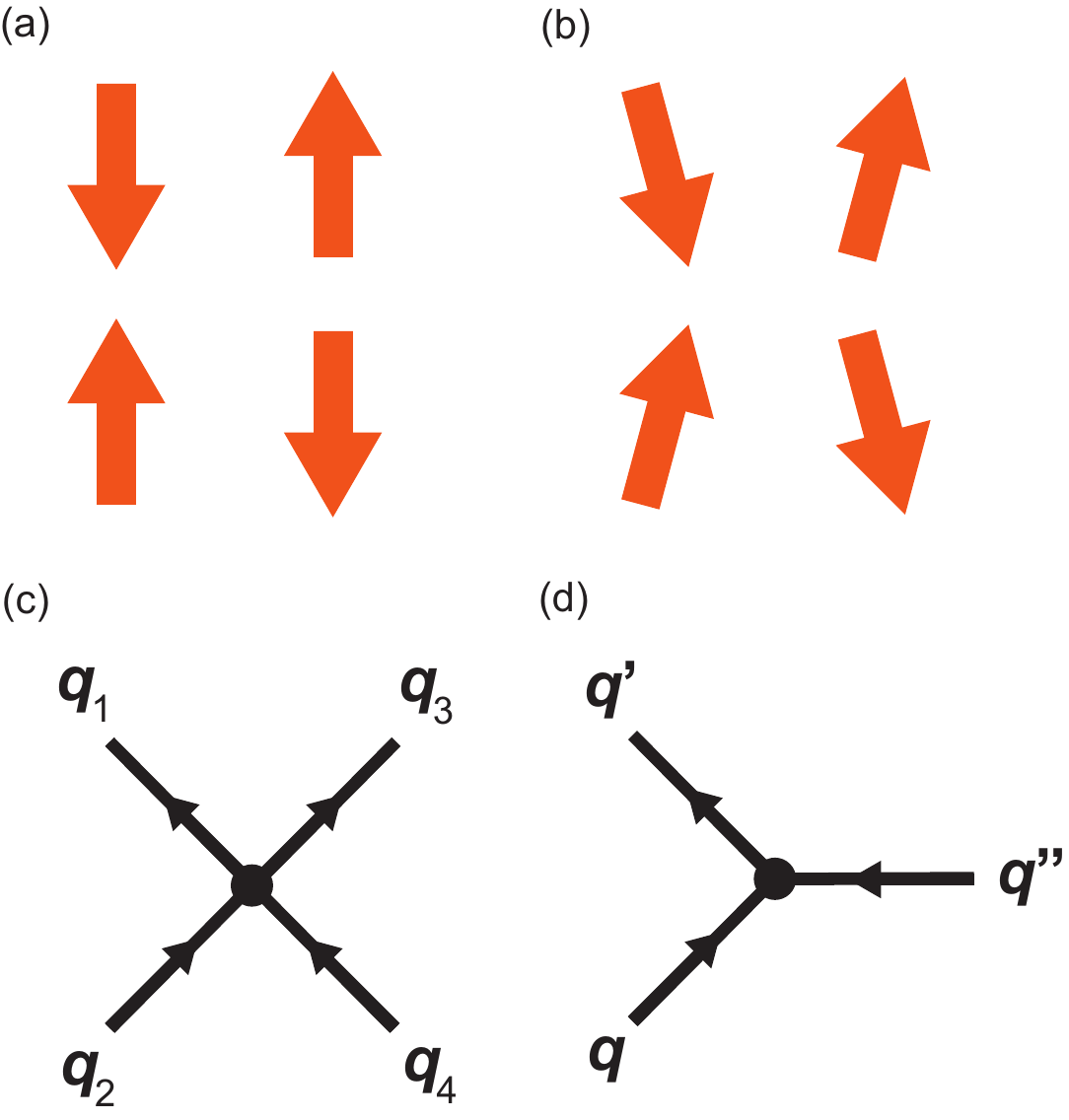}
  \caption{\label{fig1}
    Spin structures of (a) the N\'{e}el state and (b) the canted state.
    The arrows represent the magnetic moments.
    In the N\'{e}el state,
    the magnetic moments are aligned antiparallel to each other.
    Meanwhile, in the canted state, the magnetic moments are canting due to
    a magnetic field. 
    Examples of (c) the four-magnon and (d) the three-magnon scattering processes.
    The incoming and outgoing arrows represent
    the annihilations and creations of a magnon, respectively. 
    The four-magnon scattering processes
    consist of annihilation and creation processes for four magnons
    under momentum conservation $\bdq_{1}+\bdq_{3}=\bdq_{2}+\bdq_{4}$,
    whereas the three-magnon scattering processes
    consist of those for three magnons
    under momentum conservation $\bdq+\bdq^{\prime\prime}=\bdq^{\prime}$. 
    The four-magnon scattering processes are possible for both collinear and noncollinear magnets,
    whereas the three-magnon scattering processes appear only for noncollinear magnets.
  }
\end{figure}

A magnon drag is expected to be realized in noncollinear magnets,
but its possibility and effects have not been clarified yet.
Magnets are classified into 
collinear magnets and noncollinear magnets.
For collinear magnets
the magnetic moments are aligned parallel or antiparallel to each other,
whereas for noncollinear magnets
those are not.
Typical examples of collinear and noncollinear magnets
are the N\'{e}el state and a canted state, respectively
[Figs. \ref{fig1}(a) and \ref{fig1}(b)].
Spintronics or spin-caloritronics phenomena using magnons 
were initially studied
in collinear magnets~\cite{Saitoh-Nature,Saitoh-NatMat,Bauer-review,AF-SSE-theory,AF-Boltzmann},
and they have been extended to
noncollinear magnets~\cite{AF-SSE1,AF-SSE2,AF-SSE3,AF-Nakatsuji,CanAF-another1,CanAF-another2,CanAF-LLG1,CanAF-LLG2}.
Then,
there is another difference between collinear and noncollinear magnets. 
The dominant interactions between magnons
usually come from four-magnon scattering processes [Fig. \ref{fig1}(c)].
Meanwhile, 
three-magnon scattering processes [Fig. \ref{fig1}(d)],
which are described by the cubic terms in the magnon Hamiltonian,
appear only for noncollinear magnets~\cite{Cubic-Shiba,Cubic-Chubukov,Cubic-PRL}.
By analogy with the phonon drag, 
the cubic terms may cause a magnon drag that
magnons drag a magnon current.
This is distinct from
the standard magnon drag~\cite{MD-theory,MD-exp1,MD-exp2,MD-exp3}
that magnons drag an electron current;
the former works for magnetic metals and insulators,
whereas the latter works only for magnetic metals. 
Nevertheless, 
it is unclear how the cubic terms affect
magnon-transport properties of noncollinear magnets.

Here
we demonstrate that
the magnon drag induced by the cubic terms 
enhances the magnon spin current and heat current 
for a noncollinear antiferromagnet.
Our noncollinear magnet
is a three-dimensional canted antiferromagnet [Fig. \ref{fig2}(a)], such as MnF$_{2}$,
with a magnetic field along the $x$ axis.
We formulate three magnon-transport coefficients
using the linear-response theory~\cite{Kubo,Luttinger,Streda,Kontani,Ogata,NA-Ferri}
in the presence of a temperature gradient
or a nonthermal external field along the $z$ axis [Fig. \ref{fig2}(a)]:
a spin-Seebeck coefficient $S_{\textrm{m}}$,
a magnon conductivity $\sigma_{\textrm{m}}$,
and a magnon thermal conductivity $\kappa_{\textrm{m}}$.
We show that
the cubic terms lead to
the drag terms of $S_{\textrm{m}}$, $\sigma_{\textrm{m}}$, and $\kappa_{\textrm{m}}$,
which are proportional to $\tau^{2}$, the square of the magnon lifetime, 
whereas the noninteracting ones are proportional to $\tau$. 
We also show that
the drag terms cause 
low-temperature peaks of $S_{\textrm{m}}$, $\sigma_{\textrm{m}}$, and $\kappa_{\textrm{m}}$
for a strong magnetic field, at which
the cubic terms become large.
The $S_{\textrm{m}}$ obtained for a weak magnetic field
is consistent with the experiment~\cite{AF-SSE2} for MnF$_{2}$.

\section{Model}

\subsection{Magnon Hamiltonian of the canted antiferromagnet}

\begin{figure*}
  \includegraphics[width=164mm]{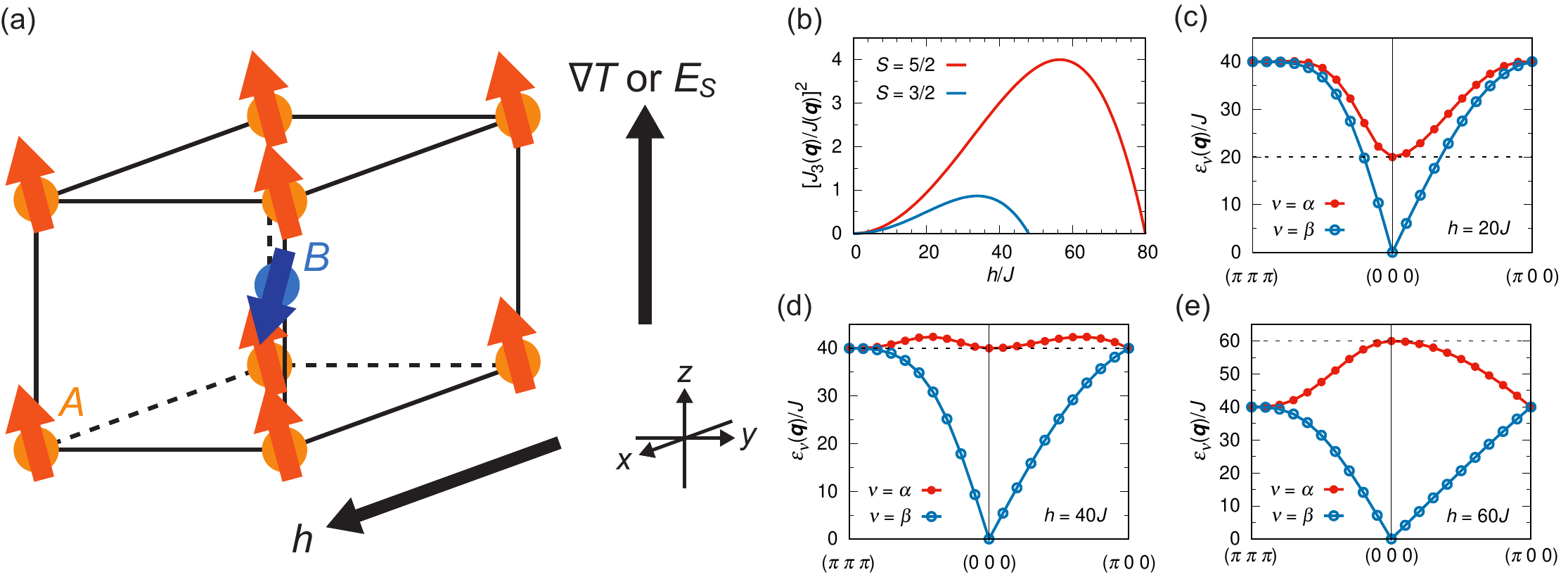}
  \caption{\label{fig2}
    (a) The spin structure of the canted antiferromagnet
    on the body-centered cubic lattice
    with the $x$, $y$, and $z$ axes.
    The case for $S=\frac{5}{2}$ corresponds to MnF$_{2}$. 
    The circles on the corners of the cube represent
    the sites on sublattice $A$,
    whereas
    that on the center represents
    the site on sublattice $B$.
    The arrows in the cube represent the canting spins.
    The magnetic field $h=-g\mu_{\textrm{B}}B$ is applied along the $x$ axis,
    where $g$ is the $g$-factor and $\mu_{\textrm{B}}$ is the Bohr magneton.
    The temperature gradient $\nabla T$
    or the non-thermal external field $\bdE_{S}$
    is applied along the $z$ axis;
    as a result,
    the magnon spin current and heat current along it are induced. 
    (b) The $h/J$ dependence of $[J_{3}(\bdq)/J(\bdq)]^{2}$
    for $S=\frac{5}{2}$ or $\frac{3}{2}$ with $\frac{N}{2}=20^{3}$ and $J=1$.
    Here $J_{3}(\bdq)$ is the coupling constant of the cubic terms.
    The red or blue curve represents that dependence for $S=\frac{5}{2}$ or $\frac{3}{2}$,
    respectively. 
    The magnon-band dispersion relations along the symmetric lines
    in the momentum space at (c) $h=20J$, (d) $40J$, and (e) $60J$ 
    for $S=\frac{5}{2}$ with $\frac{N}{2}=20^{3}$.
    The blue and red curves represent
    the energies divided by $J$ for the $\beta$-band and $\alpha$-band magnon
    [i.e., $\epsilon_{\beta}(\bdq)/J$ and $\epsilon_{\alpha}(\bdq)/J$], respectively.
    The vertical dashed lines correspond to the values of $h$. 
  }
\end{figure*}

Our noncollinear magnet is described by the spin Hamiltonian,
\begin{align}
  H=2J\sum_{\langle i,j\rangle}\bdS_{i}\cdot\bdS_{j}
  -h\sum_{i=1}^{N/2}S_{i}^{x}-h\sum_{j=1}^{N/2}S_{j}^{x}.\label{eq:Hspin}
\end{align}
Here the first term is the antiferromagnetic Heisenberg interaction
between nearest-neighbor spins,
and the others are the couplings with the magnetic field $h=-g\mu_{\textrm{B}}B$,
where $g$ and $\mu_{\textrm{B}}$ are
the $g$-factor and Bohr magneton, respectively.
We have omitted the dipolar interaction
because it may be negligible for MnF$_{2}$ (see Appendix A).
We consider a three-dimensional case on the body-centered cubic lattice [Fig. \ref{fig2}(a)];
$i$'s and $j$'s in Eq. (\ref{eq:Hspin}) are
site indices for sublattices $A$ and $B$, respectively.
In the range of $0<h < 4JzS$, where $z=8$,
the canted state for $\bdS_{i}={}^{t}(S\sin\phi\ 0\ S\cos\phi)$ and 
$\bdS_{j}={}^{t}(S\sin\phi\ 0\ -S\cos\phi)$ with $\sin\phi=\frac{h}{4JzS}$
is stabilized.
For $h=0$ or $h>4JzS=h_{\textrm{c}}$, 
the stabilized state 
becomes the N\'{e}el or the ferromagnetic state, respectively.
(Note that the energy of the canted, the N\'{e}el, or the ferromagnetic state
divided by $N/2$ is given in the mean-field approximation by
$\epsilon_{\textrm{cAF}}=-2JzS^{2}-\frac{h^{2}}{4Jz}$,
$\epsilon_{\textrm{AF}}=-2JzS^{2}$,
or $\epsilon_{\textrm{FM}}=2JzS^{2}-2Sh$, respectively.)
Therefore,
we choose the magnetic field
to be $0<h<h_{\textrm{c}}$, in the range of which
low-energy excitations can be described by magnons 
for the canted antiferromagnet.
Hereafter we set $k_{\textrm{B}}=1$, $\hbar=1$, and $a=1$,
where $a$ is the lattice constant.

To describe magnon properties,
we rewrite Eq. (\ref{eq:Hspin}) using
the Holstein-Primakoff transformation
for noncollinear
magnets~\cite{HP,Parauni1,Parauni2,Parauni3,Parauni4,NA-parauni1,NA-parauni2,NA-parauni3}.
As derived in Appendix B,
the magnon Hamiltonian of our canted antiferromagnet is written as
\begin{align}
  H=H_{0}+H_{\textrm{int}},\label{eq:H_mag}
\end{align}
where the noninteracting part $H_{0}$ consists of the quadratic terms, 
\begin{align}
  H_{0}=\sum_{\bdq}
  (a_{\bdq}^{\dagger}\ b_{\bdq}^{\dagger}\ a_{-\bdq}\ b_{-\bdq})
  \left(
  \begin{array}{@{\,}cc@{\,}}
    A_{\bdq} & B_{\bdq}\\[3pt]
    B_{\bdq} & A_{\bdq}
  \end{array}
  \right)
  \left(
  \begin{array}{@{\,}c@{\,}}
    a_{\bdq}\\[3pt]
    b_{\bdq}\\[3pt]
    a_{-\bdq}^{\dagger}\\[3pt]
    b_{-\bdq}^{\dagger}
  \end{array}
  \right),\label{eq:H0}
\end{align}
and the interaction part $H_{\textrm{int}}$ consists of the cubic terms,
\begin{align}
  H_{\textrm{int}}
  =\sum_{\bdq,\bdq^{\prime},\bdq^{\prime\prime}}
  \delta_{\bdq+\bdq^{\prime\prime},\bdq^{\prime}}J_{3}(\bdq)
  (b_{\bdq}a_{\bdq^{\prime}}^{\dagger}a_{\bdq^{\prime\prime}}-a_{\bdq}b_{\bdq^{\prime}}^{\dagger}b_{\bdq^{\prime\prime}})
  +(\textrm{H.c.}).\label{eq:Hint}
\end{align}
We have omitted the constant terms and quartic terms for simplicity.  
In Eq. (\ref{eq:H0}), 
$a_{\bdq}$ and $b_{\bdq}$ are the Fourier coefficients of the magnon operators,
the $2\times 2$ matrices $A_{\bdq}$ and $B_{\bdq}$ are given by 
$(A_{\bdq})_{11}=(A_{\bdq})_{22}=\frac{1}{2}(2Jz\cos2\phi S+h\sin\phi)=A$,
$(A_{\bdq})_{12}=(A_{\bdq})_{21}=-\frac{1}{2}\tilde{J}^{(-)}(\bdq)S=A^{\prime}(\bdq)$,
$(B_{\bdq})_{12}=(B_{\bdq})_{21}=-\frac{1}{2}\tilde{J}^{(+)}(\bdq)S=B^{\prime}(\bdq)$,
and $(B_{\bdq})_{11}=(B_{\bdq})_{22}=0$,
$\tilde{J}^{(\mp)}(\bdq)=(\cos2\phi\mp 1)J(\bdq)$,
and $J(\bdq)=8J\cos\frac{q_{x}}{2}\cos\frac{q_{y}}{2}\cos\frac{q_{z}}{2}$.
In Eq. (\ref{eq:Hint}),
\begin{align}
  J_{3}(\bdq)=\sqrt{\frac{4S}{N}}\sin2\phi J(\bdq).\label{eq:J3q}
\end{align}
Equation (\ref{eq:Hint}) is similar to that of the electron-phonon interaction
because the former and latter describe the creation and annihilation processes
for three magnons and for two electrons and a phonon, respectively.

The coupling constant of the cubic terms depends on the magnetic field
though the magnetic field dependence of the canting angle $\phi$.
Since $\sin2\phi=\frac{2h\sqrt{(4JzS)^{2}-h^{2}}}{(4JzS)^{2}}$
in our canted antiferromagnet,
$J_{3}(\bdq)$ depends on the magnetic field. 
Figure \ref{fig2}(b) shows the $h/J$ dependence of $[J_{3}(\bdq)/J(\bdq)]^{2}$
for $S=\frac{5}{2}$ or $\frac{3}{2}$.
(Note that $h_{\textrm{c}}=4JzS$ for $S=\frac{5}{2}$ or $\frac{3}{2}$
is $80J$ or $48J$, respectively.) 
We see
the coupling constant of the cubic terms for $S=\frac{5}{2}$ or $\frac{3}{2}$
is maximum at $h\sim 57J$ or $34J$, respectively.
In addition,
the coupling constant for $S=\frac{5}{2}$ at $h=65J$
is much larger than that at $h=20J$.
This suggests that
the effects of the cubic terms 
are more considerable for strong magnetic fields
than those for weak magnetic fields.
(In fact,
we will show in Sec. III B that
the cubic terms cause 
the huge enhancement of the magnon-transport coefficients 
at $h=65J$ compared with that at $h=20J$.)
We emphasize that the magnetic-field dependent coupling constant is characteristic
of canted antiferromagnets.
(Such a dependence is absent in the case of the phonon drag.)

\subsection{Noninteracting magnon bands}

We diagonalize Eq. (\ref{eq:H0}) using the Bogoliubov transformation,
\begin{align}
  \left(
  \begin{array}{@{\,}c@{\,}}
    a_{\bdq}\\[3pt]
    b_{\bdq}\\[3pt]
    a_{-\bdq}^{\dagger}\\[3pt]
    b_{-\bdq}^{\dagger}
  \end{array}
  \right)
  =
  \frac{1}{\sqrt{2}}\left(
  \begin{array}{@{\,}cccc@{\,}}
    c_{\bdq} & c_{\bdq}^{\prime} & s_{\bdq} & s_{\bdq}^{\prime}\\[3pt]
    c_{\bdq} & -c_{\bdq}^{\prime} & s_{\bdq} & -s_{\bdq}^{\prime}\\[3pt]
    s_{\bdq} & s_{\bdq}^{\prime} & c_{\bdq} & c_{\bdq}^{\prime}\\[3pt]
    s_{\bdq} & -s_{\bdq}^{\prime} & c_{\bdq} & c_{\bdq}^{\prime}
  \end{array}
  \right)
  \left(
  \begin{array}{@{\,}c@{\,}}
    \alpha_{\bdq}\\[3pt]
    \beta_{\bdq}\\[3pt]
    \alpha_{-\bdq}^{\dagger}\\[3pt]
    \beta_{-\bdq}^{\dagger}
  \end{array}
  \right),\label{eq:Bogo}
\end{align}
where $c_{\bdq}=\cosh\theta_{\bdq}$,
$s_{\bdq}=\sinh\theta_{\bdq}$,
$c^{\prime}_{\bdq}=\cosh\theta^{\prime}_{\bdq}$,
and $s^{\prime}_{\bdq}=\sinh\theta^{\prime}_{\bdq}$.
By substituting Eq. (\ref{eq:Bogo}) into Eq. (\ref{eq:H0})
and setting 
$\tanh2\theta_{\bdq}=-\frac{B^{\prime}(\bdq)}{A+A^{\prime}(\bdq)}$
and $\tanh2\theta^{\prime}_{\bdq}=\frac{B^{\prime}(\bdq)}{A-A^{\prime}(\bdq)}$,
we obtain
\begin{align}
  H_{0}=\sum_{\bdq}[\epsilon_{\alpha}(\bdq)
    \alpha_{\bdq}^{\dagger}\alpha_{\bdq}
    +\epsilon_{\beta}(\bdq)
    \beta_{\bdq}^{\dagger}\beta_{\bdq}],\label{eq:H0-band}
\end{align}
where $\epsilon_{\alpha}(\bdq)=2\sqrt{[A+A^{\prime}(\bdq)]^{2}-B^{\prime}(\bdq)^{2}}$
and $\epsilon_{\beta}(\bdq)=2\sqrt{[A-A^{\prime}(\bdq)]^{2}-B^{\prime}(\bdq)^{2}}$.
(Those choices of the hyperbolic functions are necessary
to make the off-diagonal terms zero.) 
Figures \ref{fig2}(c){--}\ref{fig2}(e) show
the magnon-band dispersion for $S=\frac{5}{2}$
at $h=20J$, $40J$, and $60J$.
The band splitting energy at $\bdq=\bdzero$ is equal to $h$ and larger than
those at the other $\bdq$'s.
This property is distinct from the property of
a two-sublattice ferrimagnet~\cite{Nakamura,AF-SSE-theory,NA-Ferri},
in which 
the band splitting energies at $\bdq=\bdzero$ and the others
are the same. 
Moreover, it indicates that
even for $T<h$,
the upper-branch magnons 
can contribute to transport properties.
(This is true, as shown in Fig. \ref{fig3}.)
Note that 
we do not study the interacting magnon-band dispersion
in this paper 
because the magnon-band energies appearing in the magnon-transport coefficients
are the noninteracting ones [see Eqs. (\ref{eq:L^0}) and (\ref{eq:L'})].

\section{Magnon-transport coefficients}

\subsection{Magnon-drag terms of $S_{\textrm{m}}$, $\sigma_{\textrm{m}}$, and $\kappa_{\textrm{m}}$}

The magnon-transport coefficients 
$S_{\textrm{m}}$, $\sigma_{\textrm{m}}$, and $\kappa_{\textrm{m}}$
are connected with $\bdj_{S}$ and $\bdj_{Q}$, 
magnon spin and heat current densities:
\begin{align}
 \left(
  \begin{array}{@{\,}c@{\,}}
    \bdj_{S}\\[3pt]
    \bdj_{Q}
  \end{array}
  \right)
  =
  \left(
  \begin{array}{@{\,}cc@{\,}}
    L_{11} & L_{12}\\[3pt]
    L_{21} & L_{22}
  \end{array}
  \right)
  \left(
  \begin{array}{@{\,}c@{\,}}
   \bdE_{S}\\[3pt]
   -\frac{\nabla T}{T}
  \end{array}
  \right),
\end{align}
where $L_{11}=\sigma_{\textrm{m}}$, $L_{12}(=L_{21})=S_{\textrm{m}}$,  
$L_{22}=\kappa_{\textrm{m}}$,
$\bdE_{S}$ is a non-thermal external field, such as a magnetic-field gradient~\cite{Nakata},
and $\nabla T$ is a temperature gradient.
(Note that our definition of $\kappa_{\textrm{m}}$
is enough to analyze its property at low temperatures
at which the magnon picture remains valid~\cite{NA-Ferri}.) 
Due to zero magnon chemical potential in equilibrium, 
$\bdj_{Q}=\bdj_{E}$ holds,
where $\bdj_{E}$ is a magnon energy current density.
By using the continuity equations,
we can express $\bdJ_{k}=N\bdj_{k}$ ($k=S,E$) 
as follows~\cite{Mahan,NA-parauni2,NA-Ferri}
(see Appendix C):
\begin{align}
  \bdJ_{k}
  =\sum_{\bdq}\sum_{l,l^{\prime}=1}^{4}
  \bdv^{k}_{ll^{\prime}}(\bdq)
  x_{\bdq l}^{\dagger}x_{\bdq l^{\prime}},\label{eq:JS,JE}
\end{align}
where
$x_{\bdq 1}=a_{\bdq}$, $x_{\bdq 2}=b_{\bdq}$,
$x_{\bdq 3}=a^{\dagger}_{-\bdq}$, $x_{\bdq 4}=b^{\dagger}_{-\bdq}$, 
$\bdv^{S}_{ll^{\prime}}(\bdq)=\bdv_{ll^{\prime}}(\bdq)=\bdv_{l^{\prime}l}(\bdq)$,
and $\bdv^{E}_{ll^{\prime}}(\bdq)=\bde_{ll^{\prime}}(\bdq)=\bde_{l^{\prime}l}(\bdq)$;
the finite terms of $\bdv_{ll^{\prime}}(\bdq)$ and $\bde_{ll^{\prime}}(\bdq)$
are given by
$\bdv_{23}(\bdq)=-\bdv_{14}(\bdq)=\frac{\partial B^{\prime}(\bdq)}{\partial \bdq}$,
$\bdv_{12}(\bdq)=-\bdv_{34}(\bdq)=\frac{\partial A^{\prime}(\bdq)}{\partial \bdq}$,
$\bde_{12}(\bdq)=-\bde_{34}(\bdq)=-2A\frac{\partial A^{\prime}(\bdq)}{\partial \bdq}$,
and 
$\bde_{11}(\bdq)=\bde_{22}(\bdq)=-\bde_{33}(\bdq)=-\bde_{44}(\bdq)=2B^{\prime}(\bdq)\frac{\partial B^{\prime}(\bdq)}{\partial \bdq}-2A^{\prime}(\bdq)\frac{\partial A^{\prime}(\bdq)}{\partial \bdq}$. 
Hereafter we concentrate on the magnon transport with $\bdE_{S}$ or
$(-\nabla T/T)$ applied along the $z$ axis [Fig. \ref{fig2}(a)].

Since the magnon lifetime $\tau$ is supposed to be long enough
to regard magnons as quasiparticles,
we derive $L_{12}$, $L_{11}$, and $L_{22}$
using the linear-response
theory~\cite{Kubo,Luttinger,Streda,Ogata,AGD,Eliashberg,NA-SCD,Kontani,NA-Ferri} 
in the limit $\tau\rightarrow \infty$.
In the linear-response theory,
$L_{\mu\eta}$ ($\mu,\eta=1,2$) is given by
\begin{align}
  L_{\mu\eta}=\lim_{\omega\rightarrow 0}
  \frac{\Phi^{\textrm{R}}_{\mu\eta}(\omega)-\Phi^{\textrm{R}}_{\mu\eta}(0)}{i\omega},\label{eq:L}
\end{align}
where 
$\Phi^{\textrm{R}}_{\mu\eta}(\omega)=\Phi_{\mu\eta}(i\Omega_{n}\rightarrow \omega+i\delta)$ ($\delta=0+$),
$\Omega_{n}=2\pi T n$ ($n>0$), 
\begin{align}
  &\Phi_{12}(i\Omega_{n})=\int_{0}^{T^{-1}}d\tau e^{i\Omega_{n}\tau}\frac{1}{N}
  \langle T_{\tau}J_{S}^{z}(\tau)J_{E}^{z}\rangle,\label{eq:Phi12}\\
  &\Phi_{11}(i\Omega_{n})=\int_{0}^{T^{-1}}d\tau e^{i\Omega_{n}\tau}\frac{1}{N}
  \langle T_{\tau}J_{S}^{z}(\tau)J_{S}^{z}\rangle,\label{eq:Phi11}\\
  &\Phi_{22}(i\Omega_{n})=\int_{0}^{T^{-1}}d\tau e^{i\Omega_{n}\tau}\frac{1}{N}
  \langle T_{\tau}J_{E}^{z}(\tau)J_{E}^{z}\rangle,\label{eq:Phi22}
\end{align}
and 
$T_{\tau}$ is the time-ordering operator~\cite{AGD}.
Since $J_{S}^{z}$ and $J_{E}^{z}$ are written as Eq. (\ref{eq:JS,JE}),
we can calculate Eqs. (\ref{eq:Phi12}){--}(\ref{eq:Phi22})
by using a method of Green's functions~\cite{AGD,Mahan,Eliashberg,NA-Ferri,Ogata}; 
in their calculations,
we treat $H_{\textrm{int}}$ in the second-order perturbation theory. 
As derived in Appendix D, 
$L_{\mu\eta}$ can be written as follows: 
\begin{align}
  L_{\mu\eta}=L_{\mu\eta}^{0}+L_{\mu\eta}^{\prime},
\end{align}
where 
$L_{\mu\eta}^{0}$ ($\mu,\eta=1,2$) is the noninteracting term,
\begin{align}
  L_{\mu\eta}^{0}=-\frac{2}{N}\sum_{\bdq}\sum_{\nu=\alpha,\beta}
  j_{\mu;\nu\nu}^{z}(\bdq)j_{\eta;\nu\nu}^{z}(\bdq)\tau
  \frac{\partial n[\epsilon_{\nu}(\bdq)]}{\partial \epsilon_{\nu}(\bdq)},\label{eq:L^0}
\end{align}
and $L_{\mu\eta}^{\prime}$ is the magnon-drag term due to the cubic terms,
\begin{align}
  L_{\mu\eta}^{\prime}=&\frac{\pi}{N^{2}}\sum_{\bdq,\bdq^{\prime}}
  \sum_{\nu,\nu^{\prime},\nu^{\prime\prime}=\alpha,\beta}
  j_{\mu;\nu\nu}^{z}(\bdq)j_{\eta;\nu^{\prime}\nu^{\prime}}^{z}(\bdq^{\prime})\tau^{2}
  \frac{\partial n[\epsilon_{\nu}(\bdq)]}{\partial \epsilon_{\nu}(\bdq)}\notag\\
  &\times S\sin^{2}2\phi \sum_{p=1,2,3}
  F^{(p)}_{\nu\nu^{\prime}\nu^{\prime\prime}}(\bdq,\bdq^{\prime}).\label{eq:L'}
\end{align}
In Eq. (\ref{eq:L^0}),
$n(x)=1/(e^{x/T}-1)$, 
$j_{1;\nu\nu}^{z}(\bdq)=v_{\nu\nu}^{z}(\bdq)$,
and $j_{2;\nu\nu}^{z}(\bdq)=e_{\nu\nu}^{z}(\bdq)$,
where
$v_{\alpha\alpha}^{z}(\bdq)=-v_{\beta\beta}^{z}(\bdq)=2v_{12}^{z}(\bdq)$,
$e_{\alpha\alpha}^{z}(\bdq)=2[e_{12}^{z}(\bdq)+e_{11}^{z}(\bdq)]$,
and $e_{\beta\beta}^{z}(\bdq)=2[-e_{12}^{z}(\bdq)+e_{11}^{z}(\bdq)]$.
In Eq. (\ref{eq:L'}), 
\begin{align}
  &F^{(2)}_{\nu\nu^{\prime}\nu^{\prime\prime}}(\bdq,\bdq^{\prime})
  =\{n[\epsilon_{\nu^{\prime\prime}}(\bdq-\bdq^{\prime})]
  -n[\epsilon_{\nu^{\prime}}(\bdq^{\prime})]\}\notag\\
  &\times 
  \delta[\epsilon_{\nu}(\bdq)-\epsilon_{\nu^{\prime}}(\bdq^{\prime})
    +\epsilon_{\nu^{\prime\prime}}(\bdq-\bdq^{\prime})]
  v^{(2)}_{\nu\nu^{\prime}\nu^{\prime\prime}}(\bdq,\bdq^{\prime}),\label{eq:F^2}\\
  &F^{(3)}_{\nu\nu^{\prime}\nu^{\prime\prime}}(\bdq,\bdq^{\prime})
  =-\{n[\epsilon_{\nu^{\prime\prime}}(\bdq-\bdq^{\prime})]
  -n[\epsilon_{\nu^{\prime}}(\bdq^{\prime})]\}\notag\\
  &\times 
  \delta[\epsilon_{\nu}(\bdq)+\epsilon_{\nu^{\prime}}(\bdq^{\prime})
    -\epsilon_{\nu^{\prime\prime}}(\bdq-\bdq^{\prime})]
  v^{(3)}_{\nu\nu^{\prime}\nu^{\prime\prime}}(\bdq,\bdq^{\prime}),\label{eq:F^3}\\
  &F^{(1)}_{\nu\nu^{\prime}\nu^{\prime\prime}}(\bdq,\bdq^{\prime})
  =\{1+n[\epsilon_{\nu^{\prime\prime}}(\bdq-\bdq^{\prime})]
  +n[\epsilon_{\nu^{\prime}}(\bdq^{\prime})]\}\notag\\
  &\times 
  \delta[\epsilon_{\nu}(\bdq)-\epsilon_{\nu^{\prime}}(\bdq^{\prime})
    -\epsilon_{\nu^{\prime\prime}}(\bdq-\bdq^{\prime})]
  v^{(1)}_{\nu\nu^{\prime}\nu^{\prime\prime}}(\bdq,\bdq^{\prime}),\label{eq:F^1}
\end{align}
and the finite components of $v^{(p)}_{\nu\nu^{\prime}\nu^{\prime\prime}}(\bdq,\bdq^{\prime})$'s 
are given by
those for $(\nu,\nu^{\prime},\nu^{\prime\prime})=(\beta,\beta,\beta)$,
$(\beta,\alpha,\alpha)$, $(\alpha,\beta,\alpha)$, 
and $(\alpha,\alpha,\beta)$ 
(for their expressions, see Appendix D). 
The most important difference between Eqs. (\ref{eq:L^0}) and (\ref{eq:L'})
is that 
$L_{\mu\eta}^{\prime}\propto \tau^{2}$, whereas $L_{\mu\eta}^{0}\propto \tau$.
This dependence is different from
that of the phonon-drag term of $L_{12}$~\cite{Ogata},
which is proportional to $\tau_{\textrm{el}}\tau_{\textrm{ph}}$,
where $\tau_{\textrm{el}}$ and $\tau_{\textrm{ph}}$ are 
the electron and phonon lifetimes, respectively.
Then, since Eq. (\ref{eq:L'}) is proportional to
the square of the coupling constant of the cubic terms,
the magnon-drag term depends on the magnetic field.
Because of this property,
our magnon drag causes unusual magnetic field dependences of the magnon-transport coefficients
(see Sec. III B).
Equation (\ref{eq:L^0}) is consistent with the expression derived
in the Boltzmann theory with the relaxation-time approximation. 

\subsection{Magnon-drag induced enhancement and low-temperature peaks
  of $S_{\textrm{m}}$, $\sigma_{\textrm{m}}$, and $\kappa_{\textrm{m}}$}

\begin{figure*}
  \includegraphics[width=180mm]{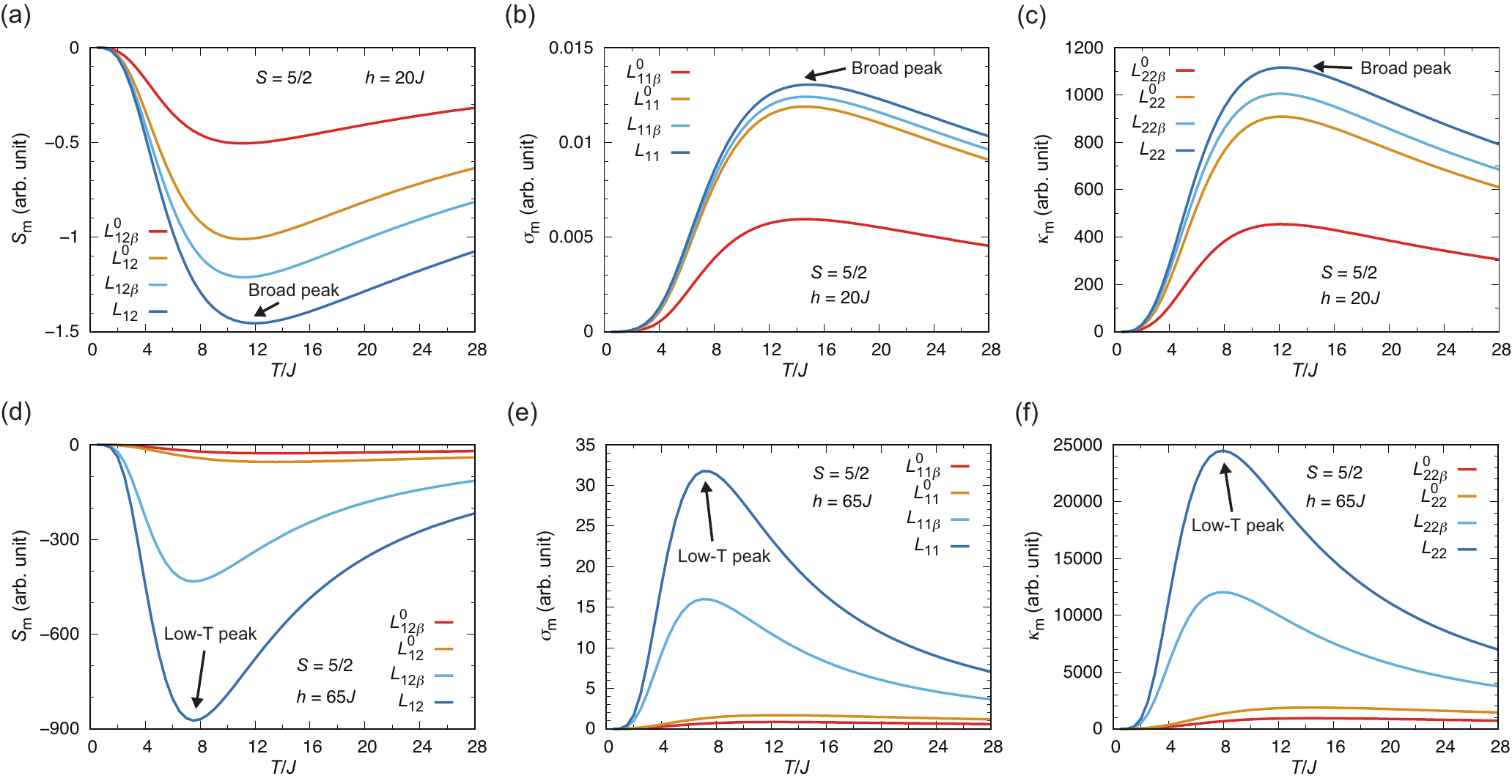}
  \caption{\label{fig3}
    The temperature dependences of (a) $S_{\textrm{m}}$,
    (b) $\sigma_{\textrm{m}}$, and (c) $\kappa_{\textrm{m}}$
    obtained in the numerical calculations for $S=\frac{5}{2}$
    with $\frac{N}{2}=20^{3}$ and $J=1$
    at a weak magnetic field $h=20J$.
    The red curves
    represent the $T/J$ dependences of $S_{\textrm{m}}=L_{12\beta}^{0}$,
    $\sigma_{\textrm{m}}=L_{11\beta}^{0}$, and $\kappa_{\textrm{m}}=L_{22\beta}^{0}$;
    the yellow curves represent 
    those of $S_{\textrm{m}}=L_{12}^{0}$,
    $\sigma_{\textrm{m}}=L_{11}^{0}$, and $\kappa_{\textrm{m}}=L_{22}^{0}$;
    the light blue curves represent
    those of $S_{\textrm{m}}=L_{12\beta}$,
    $\sigma_{\textrm{m}}=L_{11\beta}$, and $\kappa_{\textrm{m}}=L_{22\beta}$;
    and the blue curves represent 
    those of $S_{\textrm{m}}=L_{12}$,
    $\sigma_{\textrm{m}}=L_{11}$, and $\kappa_{\textrm{m}}=L_{22}$. 
    $L_{\mu\eta}^{0}=L_{\mu\eta\beta}^{0}+L_{\mu\eta\alpha}^{0}$,
    where $L_{\mu\eta\beta}^{0}$ and $L_{\mu\eta\alpha}^{0}$
    are the contributions from
    the lower-branch and higher-branch magnons
    (i.e., the $\beta$-band and $\alpha$-band magnons),
    respectively.
    $L_{\mu\eta\beta}=L_{\mu\eta}^{0}+L_{\mu\eta\beta}^{\prime}$,
    where $L_{\mu\eta\beta}^{\prime}$ is part of the drag term,
    the contribution from
    the term for $(\nu,\nu^{\prime},\nu^{\prime\prime})=(\beta,\beta,\beta)$
    in Eq. (\ref{eq:L'}).
    The temperature dependences of (d) $S_{\textrm{m}}$,
    (e) $\sigma_{\textrm{m}}$, and (f) $\kappa_{\textrm{m}}$
    obtained in the numerical calculations for $S=\frac{5}{2}$
    with $\frac{N}{2}=20^{3}$ and $J=1$
    at a strong magnetic field $h=65J$.
    The same notations as those at $h=20J$ are used. 
  }
\end{figure*}

To determine the effects of the cubic terms quantitatively, 
we evaluate $S_{\textrm{m}}$, $\sigma_{\textrm{m}}$, and $\kappa_{\textrm{m}}$ numerically.
We set $J=1$ and $S=\frac{5}{2}$.
(In the case of $S=\frac{5}{2}$,
the magnon picture for the canted antiferromagnet is valid
in the range of $0<h<80J$.)
The transition temperature
$T_{\textrm{c}}=\frac{16}{3}S(S+1)J$
is consistent with the N\'{e}el temperature $T_{\textrm{N}}$ of MnF$_{2}$ ($S=\frac{5}{2}$)
if $J\approx 1.5$ K($\approx 0.13$ meV).
Note that
$h=20J$ and $65J$ correspond to $|B|\approx 20$ and $65$ T, respectively,
using $h=-g\mu_{\textrm{B}}B$,
with $g=2$ and $J=0.13$ meV.
We belive such magnetic fields could be experimentally realized
because the magnetic field of the order of $1000T$
is experimentally accessible~\cite{HighMag}.
We perform the momentum summations 
by dividing the first Brillouin zone into a $N_{q}$-point mesh~\cite{NA-Ferri}
and setting $N_{q}=20^{3}(=N/2)$.
We consider the temperature range $0<T\leq 28J(\sim 0.6T_{\textrm{c}})$
because 
the perturbation theory with magnon-magnon interactions
can reproduce the perpendicular susceptibility of MnF$_{2}$
up to about $0.6T_{\textrm{N}}$~\cite{Kanamori}.
For simplicity,
$\tau$ is chosen to be
$\tau^{-1}=\gamma_{0}+\gamma_{1}T+\gamma_{2}T^{2}$,
where $\gamma_{0}=10^{-2}J$, $\gamma_{1}=10^{-4}$, and $\gamma_{2}=10^{-3}$.
We replace the delta functions in Eqs. (\ref{eq:F^2}){--}(\ref{eq:F^1})
by the Lorentzian ones
using
$\delta(x)\sim \frac{1}{\pi}\frac{3\gamma}{x^{2}+(3\gamma)^{2}}$,
where $\gamma=1/2\tau$. 

Figures \ref{fig3}(a){--}\ref{fig3}(c)
show the temperature dependences of
$S_{\textrm{m}}$, $\sigma_{\textrm{m}}$, and $\kappa_{\textrm{m}}$
at a weak magnetic field $h=20J$.
The contributions from the upper-branch magnons
are non-negligible even at sufficiently low temperatures
in the absence of the cubic terms
(compare the red and yellow curves of these figures).
Even in the presence of the cubic terms,
the upper-branch magnons give the non-negligible contributions
(compare the light blue and blue curves).
Furthermore,
the magnon-drag terms 
enhance $S_{\textrm{m}}$, $\sigma_{\textrm{m}}$, and $\kappa_{\textrm{m}}$.
For example, the ratios $L_{12}/L_{12}^{0}$, $L_{11}/L_{11}^{0}$,
and $L_{22}/L_{22}^{0}$ at $T=7.5J$
are about $1.4$, $1.1$, and $1.2$, respectively.
(As we will show below, these ratios become much larger for $h=65J$.)
The broad peak of $S_{\textrm{m}}$ 
is consistent with the experimental result of MnF$_{2}$~\cite{AF-SSE2}
because the voltage observed in the spin-Seebeck effect is proportional to $S_{\textrm{m}}$.

We turn to the results for a strong magnetic field $h=65J$.
Figure \ref{fig3}(d) shows that 
the magnon-drag term causes a peak at a low temperature $T=7.5J\sim 0.16 T_{\textrm{c}}$,
at which the ratio $L_{12}/L_{12}^{0}$ reaches about $22$.
This low-temperature peak is similar to that induced by the phonon drag~\cite{PD-exp,PD-theory2}.
In contrast to the phonon drag,
our magnon drag induces a low-temperature peak of $\sigma_{\textrm{m}}$,
as shown in Fig. \ref{fig3}(e).
Thus, 
our magnon drag could explain
a peak observed in $\sigma_{\textrm{m}}$~\cite{MagSpinCond}
if a noncollinear state is stabilized.
A similar peak is observed also in $\kappa_{\textrm{m}}$ [Fig. \ref{fig3}(f)].
The ratios $L_{11}/L_{11}^{0}$ and $L_{22}/L_{22}^{0}$ at $T=7.5J$
are about $23$ and $20$, respectively. 
These results suggest that 
our magnon drag can be used to
enhance the magnon spin current and heat current
by tuning the magnetic field.
The contributions from the upper-branch magnons are non-negligible also for $h=65J$ 
in the absence and presence of the cubic terms. 
Note that the larger enhancement of the magnon-transport coefficients for $h=65J$
than for $h=20J$
comes mainly from the magnetic field dependence of the coupling constant of the cubic terms.

We emphasize that 
our magnon drag can induce a similar peak
for any transport coefficient described by magnon currents.
This is an important difference between our magnon drag
and the other drag effects.
Therefore,
our magnon drag provides a mechanism 
for a low-temperature peak of a transport coefficient.

\section{Discussion}

We discuss the generality of our magnon drag. 
The mechanism for our magnon drag will work
as long as the magnon Hamiltonian contains the cubic terms.
This is because
the second-order perturbation of the cubic terms
leads to the similar magnon-drag term.
Thus, the similar enhancement of magnon-transport coefficients
may be expected to occur in other noncollinear magnets,
such as those with the Dzyaloshinsky-Moriya interaction or the dipolar interaction.
We should note that 
our magnon drag does not necessarily occur in any noncollinear magnets
because there is a noncollinear magnet in which the cubic terms are zero~\cite{NA-parauni3}.
The cubic terms in the magnon Hamiltonian are vital for our magnon drag. 

We comment on two ways to reduce the critical magnetic field 
at which a low-temperature peak appears.
One is to make $S$ smaller;
in our model for $S=3/2$,
the similar peaks of $S_{\textrm{m}}$, $\sigma_{\textrm{m}}$, and $\kappa_{\textrm{m}}$
are obtained at $h=40J$ (see Appendix E).
The other is to reduce the dimension; for example, 
in a two-dimensional canted antiferromagnet,
a low-temperature peak could be realized at smaller $h$'s.
Thus, 
the low-temperature peaks due to the magnon drag induced by the cubic terms
could be realized in various noncollinear magnets.

Our results suggest a similar drag for phonons or photons.
For example,
a phonon drag could be realized
in the presence of the anharmonicity of lattice forces,
which leads to 
the cubic terms in 
the phonon Hamiltonian~\cite{Ziman}.
Our theory is useful to study transport properties for other Bose quasiparticles.  

Finally, we discuss the differences between
the present magnon drag and another one induced by the quartic terms. 
The first-order perturbation of the quartic terms 
causes another magnon drag~\cite{NA-Ferri}.
In contrast to the present magnon drag, 
its effect is described by the drag terms proportional to $\tau$. 
Thus,
the effects of the magnon drag induced by the quartic terms are
to modify the values of the magnon-transport coefficients.
More importantly, 
it does not cause any peak, and 
its effects are negligible at low temperatures~\cite{NA-Ferri}.
Meanwhile, 
the present magnon drag causes 
the enhancement of 
$S_{\textrm{m}}$, $\sigma_{\textrm{m}}$, and $\kappa_{\textrm{m}}$
even at low temperatures
and their low-temperature peaks for the strong magnetic fields.
Since many-body effects are usually negligible at low temperatures,
the enhancement and low-temperature peaks shown in the present paper
may be unusual many-body effects.
Note that since the Holstein-Primakoff method is based on the $1/S$ expansion,
the effects of the second-order pertubation due to the cubic terms
should be compared with those of the first-order perturbation due to the quartic terms.
[The second-order terms of the cubic terms and
the first-order terms of the quartic terms are both $O(S^{0})$.]

\section{Conclusion}

In summary, we showed 
the magnon drag induced by the cubic terms.
Its effects on $S_{\textrm{m}}$, $\sigma_{\textrm{m}}$, and $\kappa_{\textrm{m}}$
are described by the terms proportional to $\tau^{2}$,
whereas the noninteracting terms are proportional to $\tau$.
Our magnon drag enhances
$S_{\textrm{m}}$, $\sigma_{\textrm{m}}$, and $\kappa_{\textrm{m}}$
even at low temperatures and 
induces their low-temperature peaks for the strong magnetic field.
It provides a mechanism for explaining
a peak observed in a transport coefficient.
The broad peak of $S_{\textrm{m}}$ for the weak magnetic field
agrees with the experimental result of MnF$_{2}$~\cite{AF-SSE2}.
Our results open a way to control the magnon spin current and heat current
of noncollinear magnets by tuning the magnetic field. 

\begin{acknowledgments}
This work was supported by JSPS KAKENHI Grants No. JP19K14664 and JP22K03532.
The author also acknowledges support from  
JST CREST Grant No. JPMJCR1901. 
\end{acknowledgments}

\appendix

\section{Estimate of the dipolar interaction energy}

We estimate the dipolar interaction energy for MnF$_{2}$.
According to the argument of Ref \onlinecite{Ash-Merm},
the dipolar interaction energy $U_{\textrm{dip}}$ will be estimated from
$U_{\textrm{dip}}\approx \frac{(g\mu_{\textrm{B}})^{2}}{r^{3}}$,
where $r$ is the distance between two magnetic dipoles.
This equation can be written as
$U_{\textrm{dip}}\approx \frac{e^{2}}{a_{0}}\frac{1}{(137)^{2}}(\frac{a_{\textrm{0}}}{r})^{3}\approx 27.2\frac{1}{(137)^{2}}(\frac{a_{\textrm{0}}}{r})^{3}$ (eV),
where $a_{\textrm{0}}\approx 0.53$ \AA.
For MnF$_{2}$,
the lattice constant along the $a$ or $b$ axis is $a\approx 4.9$ \AA,
and that along the $c$ axis is $c\approx 3.3$ \AA~\cite{MnF2-lattice}.
(This difference in the lattice constant
has been neglected in our model for simplicity.) 
Setting $r=a$ or $c$ in the above relation,
we get $U_{\textrm{dip}}\approx 1.4$ or $5.8$ $\mu$ eV, respectively.
Since these values are much smaller than the antiferromagnetic Heisenberg interaction,
the dipolar interaction may be negligible for MnF$_{2}$.

\section{Holstein-Primakoff transformation for a noncollinear magnet}

Before performing the Holstein-Primakoff transformation,
we need to rewrite the spin Hamiltonian in terms of rotated spin operators.
In general,
magnons describe spin fluctuations,
the deviations from the ground-state magnetic moments.
Since 
their directions are site-dependent in noncollinear magnets, 
we need to perform a rotation of the spin at each site~\cite{Parauni3,Parauni4,NA-parauni3}.
In our case,
the ground-state magnetic moments are characterized by
$\bdS_{i}={}^{t}(S\sin\phi\ 0\ S\cos\phi)$ 
and $\bdS_{j}={}^{t}(S\sin\phi\ 0\ -S\cos\phi)$
when 
$i$ and $j$ belong to sublattices $A$ and $B$, respectively.
Thus,
we introduce the following rotated spin operators:
\begin{align}
  &\bdS^{\prime}_{i}=R(-\phi)\bdS_{i},\label{eq:S'_i}\\
  &\bdS^{\prime}_{j}=R(\pi+\phi)\bdS_{j},\label{eq:S'_j}
\end{align}
where the rotation matrix $R(\theta)$ is given by 
$[R(\theta)]_{xx}=[R(\theta)]_{zz}=\cos\theta$,
$[R(\theta)]_{xz}=-[R(\theta)]_{zx}=\sin\theta$,
$[R(\theta)]_{yy}=1$,
and $[R(\theta)]_{xy}=[R(\theta)]_{zy}=[R(\theta)]_{yx}=[R(\theta)]_{yz}=0$.
The rotation angles have been chosen 
in order that
$\bdS^{\prime}_{i}$ and $\bdS^{\prime}_{j}$ satisfy
$\bdS^{\prime}_{i}=\bdS^{\prime}_{j}={}^{t}(0\ 0\ S)$
when 
$\bdS_{i}={}^{t}(S\sin\phi\ 0\ S\cos\phi)$
and $\bdS_{j}={}^{t}(S\sin\phi\ 0\ -S\cos\phi)$.
Because of this property,
we can apply the Holstein-Primakoff transformation similar to that for ferromagnets
to the spin Hamiltonian expressed
in terms of $\bdS^{\prime}_{i}$ and $\bdS^{\prime}_{j}$~\cite{Parauni3,Parauni4,NA-parauni3}.
Combining Eqs. (\ref{eq:S'_i}) and (\ref{eq:S'_j}) with Eq. (\ref{eq:Hspin}),
we obtain
\begin{align}
  H=&2J\sum_{\langle i,j\rangle}
  [-\cos2\phi(S_{i}^{\prime x}S_{j}^{\prime x}+S_{i}^{\prime z}S_{j}^{\prime z})
    +S_{i}^{\prime y}S_{j}^{\prime y}]\notag\\
  &+2J\sin2\phi\sum_{\langle i,j\rangle}
  (S_{i}^{\prime x}S_{j}^{\prime z}-S_{i}^{\prime z}S_{j}^{\prime x})\notag\\
  &-h\sum_{i}(\cos\phi S_{i}^{\prime x}+\sin\phi S_{i}^{\prime z})\notag\\
  &-h\sum_{j}(-\cos\phi S_{j}^{\prime x}+\sin\phi S_{j}^{\prime z}).\label{eq:Hspin'} 
\end{align}

We now apply the Holstein-Primakoff transformation,
\begin{align}
  &S_{i}^{\prime z}=S-a_{i}^{\dagger}a_{i},\
  S_{i}^{\prime +}=\sqrt{2S-a_{i}^{\dagger}a_{i}}a_{i},\
  S_{i}^{\prime -}=(S_{i}^{\prime +})^{\dagger},\label{eq:HP1}\\
  &S_{j}^{\prime z}=S-b_{j}^{\dagger}b_{j},\
  S_{j}^{\prime +}=\sqrt{2S-b_{j}^{\dagger}b_{j}}b_{j},\
  S_{j}^{\prime -}=(S_{j}^{\prime +})^{\dagger},\label{eq:HP2}
\end{align}
to Eq. (\ref{eq:Hspin'}). 
To consider magnon-magnon interactions,
we apply a $1/S$ expansion~\cite{NA-parauni3,Oguchi}
to the above equations of $S_{i}^{\prime \pm}$ and $S_{j}^{\prime \pm}$;
the result is
\begin{align}
  &S_{i}^{\prime +}\sim \sqrt{2S}a_{i}
  -\frac{1}{2\sqrt{2S}}a_{i}^{\dagger}a_{i}a_{i},\label{eq:S_i-HP1}\\
  &S_{i}^{\prime -}\sim \sqrt{2S}a_{i}^{\dagger}
  -\frac{1}{2\sqrt{2S}}a_{i}^{\dagger}a_{i}^{\dagger}a_{i},\label{eq:S_i-HP2}\\
  &S_{j}^{\prime +}\sim \sqrt{2S}b_{j}
  -\frac{1}{2\sqrt{2S}}b_{j}^{\dagger}b_{j}b_{j},\label{eq:S_j-HP1}\\
  &S_{j}^{\prime -}\sim \sqrt{2S}b_{j}^{\dagger}
  -\frac{1}{2\sqrt{2S}}b_{j}^{\dagger}b_{j}^{\dagger}b_{j}.\label{eq:S_j-HP2}
\end{align}
Substituting
Eqs. (\ref{eq:S_i-HP1}){--}(\ref{eq:S_j-HP2}) and 
the first equations of Eqs. (\ref{eq:HP1}) and (\ref{eq:HP2})
into Eq. (\ref{eq:Hspin'}),
we obtain Eq. (\ref{eq:H_mag}) with Eqs. (\ref{eq:H0}) and (\ref{eq:Hint}).

\section{Derivation of Eq. (\ref{eq:JS,JE})}

We derive Eq. (\ref{eq:JS,JE}) using the continuity equations~\cite{Mahan}.
This derivation can be performed in a way similar to
those for another noncollinear magnet~\cite{NA-parauni3} 
and for a collinear magnet~\cite{NA-Ferri}.

First, we derive the spin current operator $\bdJ_{S}$,
the $k=S$ component of Eq. (\ref{eq:JS,JE}).
We suppose that the $z$ components of $\bdS^{\prime}_{i}$ and $\bdS^{\prime}_{j}$
satisfy the following continuity equation:
\begin{align}
  \frac{d S_{m}^{\prime z}}{dt}+\nabla\cdot\bdj_{m}^{(S)}=0,\label{eq:continu-S}
\end{align}
where $\bdj_{m}^{(S)}$ is a spin current operator at site $m$.
Using Eq. (\ref{eq:continu-S}), we have
\begin{align}
  \frac{d}{dt}\Bigl(\sum_{m}\bdR_{m}S_{m}^{\prime z}\Bigr)
  =-\sum_{m}\bdR_{m}\nabla\cdot\bdj_{m}^{(S)}
  =\sum_{m}\bdj_{m}^{(S)}=\bdJ_{l}^{(S)},\label{eq:J^S-start}
\end{align}
where $l=A$ or $B$ for $m\in A$ or $B$, respectively.
(Note that $m\in A$ or $B$ means that
site $m$ belongs to sublattice $A$ or $B$, respectively.) 
In deriving this equation, we have omitted the surface contributions.
Equation (\ref{eq:J^S-start}) can be rewritten as follows:
\begin{align}
  \bdJ_{A}^{(S)}=i[H,\sum_{i}\bdR_{i}S_{i}^{\prime z}],\label{eq:J_A^S}\\
  \bdJ_{B}^{(S)}=i[H,\sum_{j}\bdR_{j}S_{j}^{\prime z}].\label{eq:J_B^S}
\end{align}
Then, the spin current operator $\bdJ_{S}$ is given by
\begin{align}
  \bdJ_{S}=\bdJ_{A}^{(S)}+\bdJ_{B}^{(S)}.\label{eq:J^S-total}
\end{align}
Since we consider the magnon system described by $H=H_{0}+H_{\textrm{int}}$, 
we replace the $H$'s, $S_{i}^{\prime z}$, and $S_{j}^{\prime z}$
in Eqs. (\ref{eq:J_A^S}) and (\ref{eq:J_B^S})
by the $H_{0}$'s, $S-a_{i}^{\dagger}a_{i}$, and $S-b_{j}^{\dagger}b_{j}$, respectively.
As a result, we have
\begin{align}
  \bdJ_{A}^{(S)}
  =\sum_{i}i\bdR_{i}[a_{i}^{\dagger}a_{i},H_{0}]
  =\sum_{i}i\bdR_{i}[a_{i}^{\dagger}a_{i},H_{AB}],\label{eq:J_A^S-2}\\
  \bdJ_{B}^{(S)}
  =\sum_{j}i\bdR_{j}[b_{j}^{\dagger}b_{j},H_{0}]
  =\sum_{j}i\bdR_{j}[b_{j}^{\dagger}b_{j},H_{AB}],\label{eq:J_B^S-2}
\end{align}
where
\begin{align}
  H_{AB}=-\tilde{J}^{(+)}S\sum_{\langle i,j\rangle}(a_{i}b_{j}+a_{i}^{\dagger}b_{j}^{\dagger})
  -\tilde{J}^{(-)}S\sum_{\langle i,j\rangle}(a_{i}b_{j}^{\dagger}+a_{i}^{\dagger}b_{j}),
\end{align}
and $\tilde{J}^{(\pm)}=(\cos2\phi\pm 1)J$. 
After some algebra,
Eqs. (\ref{eq:J_A^S-2}) and (\ref{eq:J_B^S-2}) reduce to
\begin{align}
  \bdJ_{A}^{(S)}
  =i\sum_{\langle i,j\rangle}\bdR_{i}S
  \Bigl[\tilde{J}^{(+)}(a_{i}b_{j}-a_{i}^{\dagger}b_{j}^{\dagger})
  +\tilde{J}^{(-)}(a_{i}b_{j}^{\dagger}-a_{i}^{\dagger}b_{j})\Bigr],\\
  \bdJ_{B}^{(S)}
  =i\sum_{\langle i,j\rangle}\bdR_{j}S
  \Bigl[\tilde{J}^{(+)}S(a_{i}b_{j}-a_{i}^{\dagger}b_{j}^{\dagger})
  -\tilde{J}^{(-)}(a_{i}b_{j}^{\dagger}-a_{i}^{\dagger}b_{j})\Bigr].
\end{align}
Combining these equations with Eq. (\ref{eq:J^S-total}),
we have
\begin{align}
  \bdJ_{S}
  =&i\sum_{\langle i,j\rangle}(\bdR_{i}+\bdR_{j})\tilde{J}^{(+)}S
  (a_{i}b_{j}-a_{i}^{\dagger}b_{j}^{\dagger})\notag\\
  &+i\sum_{\langle i,j\rangle}(\bdR_{i}-\bdR_{j})\tilde{J}^{(-)}S
  (a_{i}b_{j}^{\dagger}-a_{i}^{\dagger}b_{j}).\label{eq:J^S}
\end{align}
By using the Fourier coefficients of the magnon operators,
\begin{align}
  a_{i}=\sqrt{\frac{2}{N}}\sum_{\bdq}a_{\bdq}e^{-i\bdq\cdot\bdR_{i}},\
  b_{j}=\sqrt{\frac{2}{N}}\sum_{\bdq}b_{\bdq}e^{-i\bdq\cdot\bdR_{j}},\label{eq:MagOp-Fourier}
\end{align}
we can express Eq. (\ref{eq:J^S}) as follows:
\begin{align}
  \bdJ_{S}
  =&-\sum_{\bdq}\frac{\partial\tilde{J}^{(+)}(\bdq)}{\partial\bdq}S
  (a_{-\bdq}b_{\bdq}+a_{-\bdq}^{\dagger}b_{\bdq}^{\dagger})\notag\\
  &-\sum_{\bdq}\frac{\partial\tilde{J}^{(-)}(\bdq)}{\partial\bdq}S
  (a_{\bdq}b_{\bdq}^{\dagger}+a_{\bdq}^{\dagger}b_{\bdq})\notag\\
  =&-\frac{S}{2}\sum_{\bdq}\Bigl[\frac{\partial\tilde{J}^{(+)}(\bdq)}{\partial\bdq}
  (a_{-\bdq}b_{\bdq}+a_{-\bdq}^{\dagger}b_{\bdq}^{\dagger}
  -a_{\bdq}b_{-\bdq}-a_{\bdq}^{\dagger}b_{-\bdq}^{\dagger})\notag\\
  &+\frac{\partial\tilde{J}^{(-)}(\bdq)}{\partial\bdq}
  (a_{\bdq}b_{\bdq}^{\dagger}+a_{\bdq}^{\dagger}b_{\bdq}
  -a_{-\bdq}b_{-\bdq}^{\dagger}-a_{-\bdq}^{\dagger}b_{-\bdq})\Bigr],
\end{align}
where $\tilde{J}^{(\pm)}(\bdq)=(\cos2\phi\pm 1)J(\bdq)$. 
This is equivalent to the $k=S$ component of Eq. (\ref{eq:JS,JE}).

Then, we derive the energy current operator $\bdJ_{E}$,
the $k=E$ component of Eq. (\ref{eq:JS,JE}).
We suppose that
the Hamiltonian at site $m$, $h_{m}$, satisfies the following continuity equation:
\begin{align}
  \frac{d h_{m}}{dt}+\nabla\cdot\bdj_{m}^{(E)}=0,\label{eq:continu-E}
\end{align}
where 
$\bdj_{m}^{(E)}$ is an energy current operator at site $m$.
In a way similar to the derivation of $\bdJ_{S}$,
we can determine the energy current operator $\bdJ_{E}$ from
\begin{align}
  \bdJ_{E}=i[H_{0},\sum_{n}\bdR_{n}h_{n}]=i\sum_{m,n}\bdR_{n}[h_{m},h_{n}],\label{eq:JE-start}
\end{align}
where $\sum_{i=1}^{N/2}h_{i}+\sum_{j=1}^{N/2}h_{j}=H_{0}$,
$h_{i}=h_{iAA}+h_{iAB}$, and $h_{j}=h_{jBB}+h_{jBA}$.
Here 
$h_{iAA}$, $h_{iAB}$, $h_{jBB}$, and $h_{jBA}$ are given by 
\begin{align}
  &h_{iAA}=(2Jz\cos2\phi S+h\sin\phi)a_{i}^{\dagger}a_{i},\label{eq:hAA}\\
  &h_{iAB}=-\frac{1}{2}S\sum_{n}
  [\tilde{J}_{in}^{(+)}(a_{i}b_{n}+a_{i}^{\dagger}b_{n}^{\dagger})
    +\tilde{J}_{in}^{(-)}(a_{i}b_{n}^{\dagger}+a_{i}^{\dagger}b_{n})],\label{eq:hAB}\\
  &h_{jBB}=(2Jz\cos2\phi S+h\sin\phi)b_{j}^{\dagger}b_{j},\label{eq:hBB}\\
  &h_{jBA}=-\frac{1}{2}S\sum_{m}
  [\tilde{J}_{mj}^{(+)}(a_{m}b_{j}+a_{m}^{\dagger}b_{j}^{\dagger})
    +\tilde{J}_{mj}^{(-)}(a_{m}b_{j}^{\dagger}+a_{m}^{\dagger}b_{j})],\label{eq:hBA}
\end{align}
where $\tilde{J}_{ij}^{(\pm)}=(\cos2\phi\pm 1)J_{ij}$,
and $J_{ij}=J$ for nearest-neighbor $i$ and $j$. 
Combining these equations with Eq. (\ref{eq:JE-start}),
we have
\begin{align}
  \bdJ_{E}=
  &i\sum_{m,n}(\bdR_{n}-\bdR_{m})\Bigl([h_{mAA},h_{nAB}]+[h_{mAA},h_{nBA}]\notag\\
  &+[h_{mAB},h_{nBB}]+[h_{mAB},h_{nBA}]+[h_{mBB},h_{nBA}]\Bigr)\notag\\
  &+i\sum_{m,n}\bdR_{n}\Bigl([h_{mAB},h_{nAB}]+[h_{mBA},h_{nBA}]\Bigr).\label{eq:JE-next}
\end{align}
After some calculations,
Eq. (\ref{eq:JE-next}) reduces to
\begin{align}
  \bdJ_{E}=&\sum_{i,j}i(\bdR_{j}-\bdR_{i})S(2Jz\cos2\phi S+h\sin\phi)
  \tilde{J}_{ij}^{(-)}(a_{i}b_{j}^{\dagger}-a_{i}^{\dagger}b_{j})\notag\\
  &+\sum_{m,n,i}\frac{S^{2}}{2}i(\bdR_{m}-\bdR_{n})
  [\tilde{J}_{mi}^{(+)}\tilde{J}_{in}^{(+)}-\tilde{J}_{mi}^{(-)}\tilde{J}_{in}^{(-)}]
  b_{m}^{\dagger}b_{n}\notag\\
  &+\sum_{m,n,j}\frac{S^{2}}{2}i(\bdR_{m}-\bdR_{n})
  [\tilde{J}_{mj}^{(+)}\tilde{J}_{jn}^{(+)}-\tilde{J}_{mj}^{(-)}\tilde{J}_{jn}^{(-)}]
  a_{m}^{\dagger}a_{n}.\label{eq:J^E-nextnext}
\end{align}
By using the Fourier coefficients of the magnon operators [Eq. (\ref{eq:MagOp-Fourier})],
Eq. (\ref{eq:J^E-nextnext}) can be written as follows:
\begin{align}
  \bdJ_{E}
  =&\sum_{\bdq}\Bigl\{(2Jz\cos2\phi S+h\sin\phi)
  \frac{\partial \tilde{J}^{(-)}(\bdq)}{\partial\bdq}S
  (a_{\bdq}b_{\bdq}^{\dagger}+a_{\bdq}^{\dagger}b_{\bdq})\notag\\
  &+S^{2}
  \Bigl[\frac{\partial \tilde{J}^{(+)}(\bdq)}{\partial\bdq}\tilde{J}^{(+)}(\bdq)
   -\frac{\partial \tilde{J}^{(-)}(\bdq)}{\partial\bdq}\tilde{J}^{(-)}(\bdq)\Bigr]
  (a_{\bdq}^{\dagger}a_{\bdq}+b_{\bdq}^{\dagger}b_{\bdq})\Bigr\}\notag\\
  =&
  \sum_{\bdq}\frac{1}{2}(2Jz\cos2\phi S+h\sin\phi)
  \frac{\partial \tilde{J}^{(-)}(\bdq)}{\partial\bdq}S\notag\\
  &\times
  (a_{\bdq}b_{\bdq}^{\dagger}+a_{\bdq}^{\dagger}b_{\bdq}
  -a_{-\bdq}b_{-\bdq}^{\dagger}-a_{-\bdq}^{\dagger}b_{-\bdq})\notag\\
  &+\sum_{\bdq}\frac{1}{2}S^{2}
  \Bigl[\frac{\partial \tilde{J}^{(+)}(\bdq)}{\partial\bdq}\tilde{J}^{(+)}(\bdq)
    -\frac{\partial \tilde{J}^{(-)}(\bdq)}{\partial\bdq}\tilde{J}^{(-)}(\bdq)\Bigr]\notag\\
  &\times
  (a_{\bdq}^{\dagger}a_{\bdq}+b_{\bdq}^{\dagger}b_{\bdq}
  -a_{-\bdq}a_{-\bdq}^{\dagger}-b_{-\bdq}b_{-\bdq}^{\dagger}).
\end{align}
This gives the $k=E$ component of Eq. (\ref{eq:JS,JE}).

\section{Derivations of Eqs. (\ref{eq:L^0}) and (\ref{eq:L'}) 
  with the expressions of $v^{(p)}_{\nu\nu^{\prime}\nu^{\prime\prime}}(\bdq,\bdq^{\prime})$'s
  appearing in Eqs. (\ref{eq:F^2}){--}(\ref{eq:F^1})}

\begin{widetext}
\begin{figure*}
  \includegraphics[width=174mm]{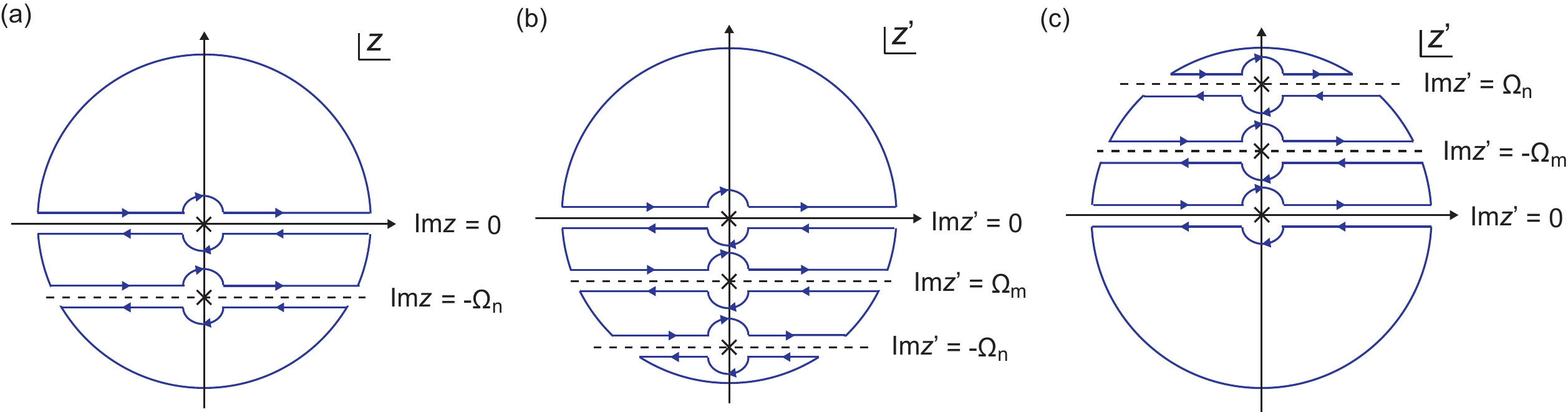}
  \caption{\label{fig4}
    The contours (a) $C$, (b) $C^{\prime}$, and (c) $C^{\prime\prime}$.
    Blue lines and curves correspond to the integral paths. 
    Crosses for $C$, $C^{\prime}$, and $C^{\prime\prime}$
    represent the poles
    at $\textrm{Im}z=0$ and $-\Omega_{n}$,
    at $\textrm{Im}z^{\prime}=0$, $\Omega_{m}$, and $-\Omega_{n}$,
    and at $\textrm{Im}z^{\prime}=0$, $-\Omega_{m}$, and $\Omega_{n}$,
    respectively.
    In these panels,
    we neglect 
    the horizontal shifts due to the noninteracting energies, such as $\epsilon_{\nu}(\bdq)$,
    for simplicity 
    because the most important information is about the imaginary parts;
    in the actual calculations,
    we consider them correctly. 
    The $C$ is used to derive Eq. (\ref{eq:Inn'});
    its contributions from the region for $-\Omega_{n}<\textrm{Im}z<0$
    are considered to replace the sums over $m$ in Eqs. (\ref{eq:tilI^a}){--}(\ref{eq:tilI^c})
    by the integrals.
    The $C^{\prime}$ or $C^{\prime\prime}$ is used to replace
    the sum over $m^{\prime}$ in Eq. (\ref{eq:tilI^a}) or (\ref{eq:tilI^b})
    or in Eq. (\ref{eq:tilI^c}), respectively.
  }
\end{figure*}

We derive Eqs. (\ref{eq:L^0}) and (\ref{eq:L'}),
$L_{\mu\eta}^{0}$ and $L_{\mu\eta}^{\prime}$ ($\mu,\eta=1,2$) in the limit $\tau\rightarrow \infty$,
and show the expressions of $v^{(p)}_{\nu\nu^{\prime}\nu^{\prime\prime}}(\bdq,\bdq^{\prime})$'s ($p=1,2,3$)
appearing in Eqs. (\ref{eq:F^2}){--}(\ref{eq:F^1}).
Since
we can derive
$L_{11}^{0}$, $L_{22}^{0}$, $L_{11}^{\prime}$, and $L_{22}^{\prime}$ 
in a way similar to the derivation of $L_{12}^{0}$ and $L_{12}^{\prime}$,
we explain the derivations of $L_{12}^{0}$ and $L_{12}^{\prime}$ below. 
Their derivations 
can be performed in a way similar to
those of the spin-Seebeck coefficient 
of a collinear magnet~\cite{NA-Ferri}
and
of the Seebeck coefficient of a metal~\cite{Ogata}.
The $v^{(p)}_{\nu\nu^{\prime}\nu^{\prime\prime}}(\bdq,\bdq^{\prime})$'s 
are given by Eqs. (\ref{eq:vbbb^1}){--}(\ref{eq:vaab^3})
with Eqs. (\ref{eq:C-S}){--}(\ref{eq:v^d3}).

First,
we derive Eq. (\ref{eq:L^0}),
the expression of $L_{\mu\eta}^{0}$ in the limit $\tau\rightarrow \infty$.
After deriving the general expression of $L_{12}^{0}$ [Eq. (\ref{eq:L12^0})],
we derive its expression in the limit $\tau\rightarrow \infty$ [Eq. (\ref{eq:L12^0-limit})].
Then, we explain how $L_{11}^{0}$ and $L_{22}^{0}$ are obtained from $L_{12}^{0}$
and show their expressions in the limit $\tau\rightarrow \infty$
[Eqs. (\ref{eq:L11^0-limit}) and (\ref{eq:L22^0-limit})]. 
Substituting Eq. (\ref{eq:JS,JE}) into Eq. (\ref{eq:Phi12}),
we have
\begin{align}
  \Phi_{12}(i\Omega_{n})
  =\frac{1}{N}\sum_{\bdq,\bdq^{\prime}}\sum_{l_{1},l_{2},l_{3},l_{4}=1}^{4}
  v_{l_{1}l_{2}}^{z}(\bdq)e_{l_{3}l_{4}}^{z}(\bdq^{\prime})
  G^{\textrm{II}}_{l_{1}l_{2}l_{3}l_{4}}(\bdq,\bdq^{\prime};i\Omega_{n}),\label{eq:Phi12-Ap}
\end{align}
where
\begin{align}
  G^{\textrm{II}}_{l_{1}l_{2}l_{3}l_{4}}(\bdq,\bdq^{\prime};i\Omega_{n})
  =\int_{0}^{T^{-1}}d\tau e^{i\Omega_{n}\tau}
  \langle T_{\tau}x_{\bdq l_{1}}^{\dagger}(\tau)x_{\bdq l_{2}}(\tau)
  x_{\bdq^{\prime}l_{3}}^{\dagger}x_{\bdq^{\prime}l_{4}}\rangle.\label{eq:G^II}
\end{align}
The expectation value in Eq. (\ref{eq:G^II}) can be calculated
by using the method of Green's functions~\cite{AGD,Mahan,Eliashberg}. 
Equation (\ref{eq:Phi12-Ap}) provides a starting point
to derive $L_{12}^{0}$ and $L_{12}^{\prime}$.
To derive $L_{12}^{0}$,
we evaluate Eq. (\ref{eq:G^II}) without the effects of $H_{\textrm{int}}$
using the Wick's theorem; the result is
\begin{align}
  G^{\textrm{II}(0)}_{l_{1}l_{2}l_{3}l_{4}}(\bdq,\bdq^{\prime};i\Omega_{n})
  =\delta_{\bdq,\bdq^{\prime}}
  T\sum_{m}G_{l_{4}l_{1}}(\bdq,i\Omega_{m})G_{l_{2}l_{3}}(\bdq,i\Omega_{n+m}),\label{eq:G^II0}
\end{align}
where $G_{ll^{\prime}}(\bdq,i\Omega_{m})$
is the magnon Green's function in the Matsubara-frequency representation,
\begin{align}
  G_{ll^{\prime}}(\bdq,i\Omega_{m})
  =\int_{0}^{T^{-1}}d\tau e^{i\Omega_{m}\tau}G_{ll^{\prime}}(\bdq,\tau)
  =-\int_{0}^{T^{-1}}d\tau e^{i\Omega_{m}\tau}
  \langle T_{\tau}x_{\bdq l}(\tau)x^{\dagger}_{\bdq l^{\prime}}\rangle,\label{eq:G_ll'}
\end{align}
and $\Omega_{m}=2\pi T m$.
Substituting Eq. (\ref{eq:G^II0}) into Eq. (\ref{eq:Phi12-Ap}),
we obtain
\begin{align}
  \Phi_{12}^{(0)}(i\Omega_{n})
  =\frac{1}{N}\sum_{\bdq}\sum_{l_{1},l_{2},l_{3},l_{4}=1}^{4}
  v_{l_{1}l_{2}}^{z}(\bdq)e_{l_{3}l_{4}}^{z}(\bdq)
  T\sum_{m}G_{l_{4}l_{1}}(\bdq,i\Omega_{m})G_{l_{2}l_{3}}(\bdq,i\Omega_{n+m}).\label{eq:Phi12^0}
\end{align}
By using the Bogoliubov transformation [Eq. (\ref{eq:Bogo})],
\begin{align}
  x_{\bdq l}=\sum_{\nu=\alpha_{1},\beta_{1},\alpha_{2},\beta_{2}}(P_{\bdq})_{l\nu}x^{\prime}_{\bdq \nu},\label{eq:Bogo-Ap}
\end{align}
where
\begin{align}
  &x^{\prime}_{\bdq \alpha_{1}}=\alpha_{\bdq},\
  x^{\prime}_{\bdq \beta_{1}}=\beta_{\bdq},\ 
  x^{\prime}_{\bdq \alpha_{2}}=\alpha^{\dagger}_{-\bdq},\
  x^{\prime}_{\bdq \beta_{2}}=\beta^{\dagger}_{-\bdq},\\
  &(P_{\bdq})_{1\alpha_{1}}=(P_{\bdq})_{2\alpha_{1}}
  =(P_{\bdq})_{3\alpha_{2}}=(P_{\bdq})_{4\alpha_{2}}
  =\frac{1}{\sqrt{2}}\cosh\theta_{\bdq},\label{eq:P-1}\\
  &(P_{\bdq})_{3\alpha_{1}}=(P_{\bdq})_{4\alpha_{1}}
  =(P_{\bdq})_{1\alpha_{2}}=(P_{\bdq})_{2\alpha_{2}}
  =\frac{1}{\sqrt{2}}\sinh\theta_{\bdq},\label{eq:P-2}\\
  &(P_{\bdq})_{1\beta_{1}}=-(P_{\bdq})_{2\beta_{1}}
  =(P_{\bdq})_{3\beta_{2}}=-(P_{\bdq})_{4\beta_{2}}
  =\frac{1}{\sqrt{2}}\cosh\theta^{\prime}_{\bdq},\label{eq:P-3}\\
  &(P_{\bdq})_{3\beta_{1}}=-(P_{\bdq})_{4\beta_{1}}
  =(P_{\bdq})_{1\beta_{2}}=-(P_{\bdq})_{2\beta_{2}}
  =\frac{1}{\sqrt{2}}\sinh\theta^{\prime}_{\bdq},\label{eq:P-4}
\end{align}
we can rewrite Eq. (\ref{eq:Phi12^0}) as follows:
\begin{align}
  \Phi_{12}^{(0)}(i\Omega_{n})
  =\frac{1}{N}\sum_{\bdq}\sum_{\nu,\nu^{\prime}=\alpha_{1},\beta_{1},\alpha_{2},\beta_{2}}
  v_{\nu^{\prime}\nu}^{z}(\bdq)e_{\nu\nu^{\prime}}^{z}(\bdq)
  T\sum_{m}G_{\nu^{\prime}}(\bdq,i\Omega_{m})G_{\nu}(\bdq,i\Omega_{n+m}),\label{eq:Phi12^0-rewrite}
\end{align}
where
\begin{align}
  &v_{\nu^{\prime}\nu}^{z}(\bdq)=\sum_{l_{1},l_{2}=1}^{4}
  (P_{\bdq})_{l_{1}\nu^{\prime}}(P_{\bdq})_{l_{2}\nu}v_{l_{1}l_{2}}^{z}(\bdq),\label{eq:v}\\
  &e_{\nu\nu^{\prime}}^{z}(\bdq)=\sum_{l_{3},l_{4}=1}^{4}
  (P_{\bdq})_{l_{3}\nu}(P_{\bdq})_{l_{4}\nu^{\prime}}e_{l_{3}l_{4}}^{z}(\bdq),\label{eq:e}\\
  &G_{\alpha_{1}}(\bdq,i\Omega_{m})=\frac{1}{i\Omega_{m}-\epsilon_{\alpha}(\bdq)},\
  G_{\beta_{1}}(\bdq,i\Omega_{m})=\frac{1}{i\Omega_{m}-\epsilon_{\beta}(\bdq)},\label{eq:G1}\\
  &G_{\alpha_{2}}(\bdq,i\Omega_{m})=-\frac{1}{i\Omega_{m}+\epsilon_{\alpha}(\bdq)},\
  G_{\beta_{2}}(\bdq,i\Omega_{m})=-\frac{1}{i\Omega_{m}+\epsilon_{\beta}(\bdq)}.\label{eq:G2}
\end{align}
Then,
to perform the analytic continuation, 
we replace 
the Matsubara-frequency summation in Eq. (\ref{eq:Phi12^0-rewrite})
by the corresponding integral~\cite{Eliashberg,NA-Ferri};
the result is
\begin{align}
  T\sum_{m}G_{\nu^{\prime}}(\bdq,i\Omega_{m})G_{\nu}(\bdq,i\Omega_{n+m})
  &=\int_{C}\frac{dz}{2\pi i}n(z)
  G_{\nu^{\prime}}(\bdq,z)G_{\nu}(\bdq,z+i\Omega_{n})
  +T[G_{\nu^{\prime}}(\bdq,0)G_{\nu}(\bdq,i\Omega_{n})
    +G_{\nu^{\prime}}(\bdq,-i\Omega_{n})G_{\nu}(\bdq,0)]\notag\\
  &=\int_{-\infty}^{\infty}\frac{dz}{2\pi i}n(z)
  \Bigl\{
  G_{\nu}^{(\textrm{R})}(\bdq,z+i\Omega_{n})
  [G_{\nu^{\prime}}^{(\textrm{R})}(\bdq,z)-G_{\nu^{\prime}}^{(\textrm{A})}(\bdq,z)]\notag\\
  &\ \ \ \ \ \ \ \ \ \ \ \ \ \ \ \ \ \ \ \ \ 
  +[G_{\nu}^{(\textrm{R})}(\bdq,z)-G_{\nu}^{(\textrm{A})}(\bdq,z)]
  G_{\nu^{\prime}}^{(\textrm{A})}(\bdq,z-i\Omega_{n})
  \Bigr\},\label{eq:Inn'}
\end{align}
where the contour $C$ is shown in Fig. \ref{fig4}(a),
$n(z)$ is the Bose distribution function $n(z)=1/(e^{z/T}-1)$,
$G_{\nu}^{(\textrm{R})}(\bdq,z)$ and
$G_{\nu}^{(\textrm{A})}(\bdq,z)[=G_{\nu}^{(\textrm{R})}(\bdq,z)^{\ast}]$
are the retarded and advanced magnon Green's functions, respectively, 
\begin{align}
  &G_{\alpha_{1}}^{(\textrm{R})}(\bdq,z)=\frac{1}{z+i\gamma-\epsilon_{\alpha}(\bdq)},\
  G_{\beta_{1}}^{(\textrm{R})}(\bdq,z)=\frac{1}{z+i\gamma-\epsilon_{\beta}(\bdq)},\label{eq:G^R-1}\\
  &G_{\alpha_{2}}^{(\textrm{R})}(\bdq,z)=-\frac{1}{z+i\gamma+\epsilon_{\alpha}(\bdq)},\
  G_{\beta_{2}}^{(\textrm{R})}(\bdq,z)=-\frac{1}{z+i\gamma+\epsilon_{\beta}(\bdq)},\label{eq:G^R-2}
\end{align}
and $\gamma(=1/2\tau)$ is the magnon damping.
By combining Eq. (\ref{eq:Inn'}) with Eq. (\ref{eq:Phi12^0-rewrite})
and performing the analytic continuation $i\Omega_{n}\rightarrow \omega+i\delta$
[i.e., $\Phi^{\textrm{R}(0)}_{12}(\omega)=\Phi_{12}^{(0)}(i\Omega_{n}\rightarrow \omega+i\delta)$],
we obtain
\begin{align}
  \Phi^{\textrm{R}(0)}_{12}(\omega)
  =&\frac{1}{N}\sum_{\bdq}\sum_{\nu,\nu^{\prime}=\alpha_{1},\beta_{1},\alpha_{2},\beta_{2}}
  v_{\nu^{\prime}\nu}^{z}(\bdq)e_{\nu\nu^{\prime}}^{z}(\bdq)
  \int_{-\infty}^{\infty}\frac{dz}{2\pi i}n(z)\notag\\
  &\times\Bigl\{
  G_{\nu}^{(\textrm{R})}(\bdq,z+\omega)
  [G_{\nu^{\prime}}^{(\textrm{R})}(\bdq,z)-G_{\nu^{\prime}}^{(\textrm{A})}(\bdq,z)]
  +[G_{\nu}^{(\textrm{R})}(\bdq,z)-G_{\nu}^{(\textrm{A})}(\bdq,z)]
  G_{\nu^{\prime}}^{(\textrm{A})}(\bdq,z-\omega)
  \Bigr\}.\label{eq:Phi12^R0}
\end{align}
After some calculations, 
Eq. (\ref{eq:Phi12^R0}) reduces to
\begin{align}
  \Phi^{\textrm{R}(0)}_{12}(\omega)
  \sim \Phi^{\textrm{R}(0)}_{12}(0)
  -\frac{\omega}{2N}\sum_{\bdq}\sum_{\nu,\nu^{\prime}=\alpha_{1},\beta_{1},\alpha_{2},\beta_{2}}
  v_{\nu^{\prime}\nu}^{z}(\bdq)e_{\nu\nu^{\prime}}^{z}(\bdq)
  \int_{-\infty}^{\infty}\frac{dz}{2\pi i}\frac{\partial n(z)}{\partial z}
      [-4\textrm{Im}G_{\nu}^{(\textrm{R})}(\bdq,z)\textrm{Im}G_{\nu^{\prime}}^{(\textrm{R})}(\bdq,z)].
      \label{eq:Phi12^R0-last}
\end{align}
In deriving this equation,
we have used 
$f(z\pm\omega)=f(z)\pm\omega\frac{\partial f(z)}{\partial z}+O(\omega^{2})$,
$v_{\nu^{\prime}\nu}^{z}(\bdq)=v_{\nu\nu^{\prime}}^{z}(\bdq)$,
and $e_{\nu\nu^{\prime}}^{z}(\bdq)=e_{\nu^{\prime}\nu}^{z}(\bdq)$.
Combining Eq. (\ref{eq:Phi12^R0-last}) with Eq. (\ref{eq:L}),
we have
\begin{align}
  \hspace{-8pt}
  L_{12}^{0}
  =\lim_{\omega\rightarrow 0}
  \frac{\Phi^{\textrm{R}(0)}_{12}(\omega)-\Phi^{\textrm{R}(0)}_{12}(0)}{i\omega}
  =-\frac{1}{N}\sum_{\bdq}\sum_{\nu,\nu^{\prime}=\alpha_{1},\beta_{1},\alpha_{2},\beta_{2}}
  v_{\nu^{\prime}\nu}^{z}(\bdq)e_{\nu\nu^{\prime}}^{z}(\bdq)
  \int_{-\infty}^{\infty}\frac{dz}{\pi}\frac{\partial n(z)}{\partial z}
  \textrm{Im}G_{\nu}^{(\textrm{R})}(\bdq,z)\textrm{Im}G_{\nu^{\prime}}^{(\textrm{R})}(\bdq,z).\label{eq:L12^0}
\end{align}
Then, we take the limit $\tau=1/2\gamma\rightarrow \infty$.
In this limit,
the integral part in Eq. (\ref{eq:L12^0}) reduces to
\begin{align}
  I_{\nu\nu^{\prime}}(\bdq)=\int_{-\infty}^{\infty}\frac{dz}{\pi}\frac{\partial n(z)}{\partial z}
  \textrm{Im}G_{\nu}^{(\textrm{R})}(\bdq,z)\textrm{Im}G_{\nu^{\prime}}^{(\textrm{R})}(\bdq,z)
  \sim
  \begin{cases}
    \dfrac{1}{2\gamma}\dfrac{\partial n[\epsilon_{\alpha}(\bdq)]}{\partial \epsilon_{\alpha}(\bdq)}
    \ \ (\nu=\nu^{\prime}=\alpha_{1},\alpha_{2})\\
    \dfrac{1}{2\gamma}\dfrac{\partial n[\epsilon_{\beta}(\bdq)]}{\partial \epsilon_{\beta}(\bdq)}
    \ \ (\nu=\nu^{\prime}=\beta_{1},\beta_{2})\\
    0 \ \ \ \ \ \ \ \ \ \ \ \ \ \ \ \ \ (\nu\neq\nu^{\prime})
  \end{cases}.\label{eq:I_nn'}  
\end{align}
These limiting expressions can be obtained by using Eqs. (\ref{eq:G^R-1}) and (\ref{eq:G^R-2})
and doing the integral~\cite{NA-Ferri}.
Combining Eq. (\ref{eq:I_nn'}) with Eq. (\ref{eq:L12^0}), 
we obtain the expression of $L_{12}^{0}$ in the limit
$\tau=1/2\gamma\rightarrow \infty$,
\begin{align}
  L_{12}^{0}\sim -\frac{1}{N}\sum_{\bdq}\sum_{\nu=\alpha_{1},\beta_{1},\alpha_{2},\beta_{2}}
  v_{\nu\nu}^{z}(\bdq)e_{\nu\nu}^{z}(\bdq)
  \tau\frac{\partial n[\epsilon_{\nu}(\bdq)]}{\partial \epsilon_{\nu}(\bdq)},
  \label{eq:L12^0-rewrite}
\end{align}
where 
$\epsilon_{\alpha_{1}}(\bdq)=\epsilon_{\alpha_{2}}(\bdq)=\epsilon_{\alpha}(\bdq)$
and $\epsilon_{\beta_{1}}(\bdq)=\epsilon_{\beta_{2}}(\bdq)=\epsilon_{\beta}(\bdq)$.
In addition,
using Eqs. (\ref{eq:v}) and (\ref{eq:e}) and Eqs. (\ref{eq:P-1}){--}(\ref{eq:P-4}),
we have
\begin{align}
  &v_{\alpha_{1}\alpha_{1}}^{z}(\bdq)=-v_{\beta_{1}\beta_{1}}^{z}(\bdq)
  =-v_{\alpha_{2}\alpha_{2}}^{z}(\bdq)=v_{\beta_{2}\beta_{2}}^{z}(\bdq)
  =2v_{12}^{z}(\bdq),\label{eq:v-band}\\
  &e_{\alpha_{1}\alpha_{1}}^{z}(\bdq)=-e_{\alpha_{2}\alpha_{2}}^{z}(\bdq)
  =2[e_{12}^{z}(\bdq)+e_{11}^{z}(\bdq)],\label{eq:e-band1}\\
  &e_{\beta_{1}\beta_{1}}^{z}(\bdq)=-e_{\beta_{2}\beta_{2}}^{z}(\bdq)
  =2[-e_{12}^{z}(\bdq)+e_{11}^{z}(\bdq)],\label{eq:e-band2}
\end{align}
where $v_{12}^{z}(\bdq)$, $e_{12}^{z}(\bdq)$, and $e_{11}^{z}(\bdq)$ are defined
below Eq. (\ref{eq:JS,JE}).
Thus, Eq. (\ref{eq:L12^0-rewrite}) reduces to
\begin{align}
  L_{12}^{0}\sim -\frac{2}{N}\sum_{\bdq}\sum_{\nu=\alpha,\beta}
  v_{\nu\nu}^{z}(\bdq)e_{\nu\nu}^{z}(\bdq)
  \tau\frac{\partial n[\epsilon_{\nu}(\bdq)]}{\partial \epsilon_{\nu}(\bdq)},
  \label{eq:L12^0-limit}
\end{align}
where $v_{\alpha\alpha}^{z}(\bdq)=-v_{\beta\beta}^{z}(\bdq)=v_{\alpha_{1}\alpha_{1}}^{z}(\bdq)$,
$e_{\alpha\alpha}^{z}(\bdq)=e_{\alpha_{1}\alpha_{1}}^{z}(\bdq)$,
and $e_{\beta\beta}^{z}(\bdq)=e_{\beta_{1}\beta_{1}}^{z}(\bdq)$.
Then,
Eqs. (\ref{eq:JS,JE}) and (\ref{eq:Phi12}){--}(\ref{eq:Phi22}) 
show that
$L_{11}^{0}$ and $L_{22}^{0}$ are obtained
by replacing $e_{\nu\nu}^{z}(\bdq)$ in Eq. (\ref{eq:L12^0-limit})
by $v_{\nu\nu}^{z}(\bdq)$
and by replacing $v_{\nu\nu}^{z}(\bdq)$ in Eq. (\ref{eq:L12^0-limit})
by $e_{\nu\nu}^{z}(\bdq)$, respectively.
Therefore, $L_{11}^{0}$ and $L_{22}^{0}$ in the limit $\tau\rightarrow \infty$ are given by
\begin{align}
  L_{11}^{0}&\sim -\frac{2}{N}\sum_{\bdq}\sum_{\nu=\alpha,\beta}
  v_{\nu\nu}^{z}(\bdq)v_{\nu\nu}^{z}(\bdq)
  \tau\frac{\partial n[\epsilon_{\nu}(\bdq)]}{\partial \epsilon_{\nu}(\bdq)},
  \label{eq:L11^0-limit}\\
  L_{22}^{0}&\sim -\frac{2}{N}\sum_{\bdq}\sum_{\nu=\alpha,\beta}
  e_{\nu\nu}^{z}(\bdq)e_{\nu\nu}^{z}(\bdq)
  \tau\frac{\partial n[\epsilon_{\nu}(\bdq)]}{\partial \epsilon_{\nu}(\bdq)}.
  \label{eq:L22^0-limit}
\end{align}
Equations (\ref{eq:L12^0-limit}){--}(\ref{eq:L22^0-limit})
give Eq. (\ref{eq:L^0}). 

Next, we derive Eq. (\ref{eq:L'}),
the expression of $L_{\mu\eta}^{\prime}$
in the limit $\tau\rightarrow \infty$
[Eqs. (\ref{eq:L12'-last}), (\ref{eq:L11'-last}), and (\ref{eq:L22'-last})],
and show
the explicit expressions of $v^{(p)}_{\nu\nu^{\prime}\nu^{\prime\prime}}(\bdq,\bdq^{\prime})$'s
[Eqs. (\ref{eq:vbbb^1}){--}(\ref{eq:vaab^3})].
(This derivation can be done in a way similar to
that of the phonon-drag term of a metal~\cite{Ogata}.)
Before evaluating Eq. (\ref{eq:G^II}) with the effects of $H_{\textrm{int}}$,
we express $H_{\textrm{int}}$ in terms of the operators $x_{\bdq l}$ and $x_{\bdq l}^{\dagger}$.
Since $H_{\textrm{int}}$ is defined as Eq. (\ref{eq:Hint}), 
we have
\begin{align}
  H_{\textrm{int}}
  =&\frac{1}{2}\sum_{\bdq,\bdq^{\prime},\bdq^{\prime\prime}}
  \delta_{\bdq+\bdq^{\prime\prime},\bdq^{\prime}}J_{3}(\bdq)
  (b_{\bdq}a_{\bdq^{\prime}}^{\dagger}a_{\bdq^{\prime\prime}}
   -a_{\bdq}b_{\bdq^{\prime}}^{\dagger}b_{\bdq^{\prime\prime}}
   +b_{-\bdq}a_{-\bdq^{\prime}}^{\dagger}a_{-\bdq^{\prime\prime}}
   -a_{-\bdq}b_{-\bdq^{\prime}}^{\dagger}b_{-\bdq^{\prime\prime}})
  +(\textrm{H.c.})\notag\\
  =&\frac{1}{2}\sum_{\bdq,\bdq^{\prime},\bdq^{\prime\prime}}
  \delta_{\bdq+\bdq^{\prime\prime},\bdq^{\prime}}J_{3}(\bdq)
  \Bigl[\sum_{l=1}^{2}\textrm{sgn}(l)
  (x_{\bdq l}x_{\bdq^{\prime} \bar{l}}^{\dagger}x_{\bdq^{\prime\prime} \bar{l}}
  +x_{\bdq l}^{\dagger}x_{\bdq^{\prime\prime} \bar{l}}^{\dagger}x_{\bdq^{\prime} \bar{l}})
  +\sum_{l=3}^{4}\textrm{sgn}(l)
  (x_{\bdq l}x_{\bdq^{\prime\prime} \bar{l}}x_{\bdq^{\prime} \bar{l}}^{\dagger}
  +x_{\bdq l}^{\dagger}x_{\bdq^{\prime} \bar{l}}x_{\bdq^{\prime\prime} \bar{l}}^{\dagger})\Bigr],\label{eq:Hint-Ap}
\end{align}
where
\begin{align}
  \textrm{sgn}(l)=
  \begin{cases}
    -1\ \ \ (l=1,3)\\
    1\ \ \ \ \ (l=2,4)
  \end{cases},\
  \bar{l}=
  \begin{cases}
    2\ \ \ (l=1)\\
    1\ \ \ (l=2)\\
    4\ \ \ (l=3)\\
    3\ \ \ (l=4)
  \end{cases}.
\end{align}
To derive $L_{12}^{\prime}$,
we evaluate Eq. (\ref{eq:G^II}) in the second-order perturbation theory~\cite{AGD,Mahan}
using the Wick's theorem and Eqs. (\ref{eq:Bogo-Ap}) and (\ref{eq:Hint-Ap});
the result is
\begin{align}
  \Delta G^{\textrm{II}}_{l_{1}l_{2}l_{3}l_{4}}(\bdq,\bdq^{\prime};i\Omega_{n})
  =&\int_{0}^{T^{-1}}d\tau e^{i\Omega_{n}\tau}\int_{0}^{T^{-1}}d\tau_{1}\int_{0}^{T^{-1}}d\tau_{2}
  \frac{1}{2}\langle T_{\tau}x_{\bdq l_{1}}^{\dagger}(\tau)x_{\bdq l_{2}}(\tau)
  x_{\bdq^{\prime}l_{3}}^{\dagger}x_{\bdq^{\prime}l_{4}}
  H_{\textrm{int}}(\tau_{1})H_{\textrm{int}}(\tau_{2})\rangle\notag\\
  =&\int_{0}^{T^{-1}}d\tau e^{i\Omega_{n}\tau}\int_{0}^{T^{-1}}d\tau_{1}\int_{0}^{T^{-1}}d\tau_{2}
  \sum_{\nu_{1},\nu_{2},\nu_{3},\nu_{4},\nu_{5}=\alpha_{1},\beta_{1},\alpha_{2},\beta_{2}}
  (P_{\bdq})_{l_{1}\nu_{1}}(P_{\bdq})_{l_{2}\nu_{2}}(P_{\bdq^{\prime}})_{l_{3}\nu_{3}}(P_{\bdq^{\prime}})_{l_{4}\nu_{4}}\notag\\
  &\times 
  \sum_{k=a,b,c}
  \tilde{V}_{\nu_{1}\nu_{2}\nu_{3}\nu_{4}\nu_{5}}^{(k)}(\bdq,\bdq^{\prime})
  f_{\nu_{1}\nu_{2}\nu_{3}\nu_{4}\nu_{5}}^{(k)}(\bdq,\bdq^{\prime};\tau,\tau_{1},\tau_{2}),
  \label{eq:DelG^II}
\end{align}
where
\begin{align}
  \tilde{V}_{\nu_{1}\nu_{2}\nu_{3}\nu_{4}\nu_{5}}^{(a)}(\bdq,\bdq^{\prime})
  =-\frac{1}{4}\sum_{l,l^{\prime}=1}^{4}\textrm{sgn}(l)\textrm{sgn}(l^{\prime})
  &[J_{3}(\bdq)^{2}(P_{\bdq})_{l\nu_{1}}(P_{\bdq})_{l^{\prime}\nu_{2}}
   (P_{\bdq^{\prime}})_{\bar{l^{\prime}}\nu_{3}}(P_{\bdq^{\prime}})_{\bar{l}\nu_{4}}
   (P_{\bdq^{\prime}-\bdq})_{\bar{l}\nu_{5}}(P_{\bdq^{\prime}-\bdq})_{\bar{l^{\prime}}\nu_{5}}\notag\\[-11pt]
  &+J_{3}(\bdq)J_{3}(\bdq^{\prime}-\bdq)(P_{\bdq})_{l\nu_{1}}(P_{\bdq})_{\bar{l^{\prime}}\nu_{2}}
   (P_{\bdq^{\prime}})_{\bar{l^{\prime}}\nu_{3}}(P_{\bdq^{\prime}})_{\bar{l}\nu_{4}}
   (P_{\bdq^{\prime}-\bdq})_{\bar{l}\nu_{5}}(P_{\bdq^{\prime}-\bdq})_{l^{\prime}\nu_{5}}\notag\\
  &+J_{3}(\bdq^{\prime}-\bdq)J_{3}(\bdq)(P_{\bdq})_{\bar{l}\nu_{1}}(P_{\bdq})_{l^{\prime}\nu_{2}}
   (P_{\bdq^{\prime}})_{\bar{l^{\prime}}\nu_{3}}(P_{\bdq^{\prime}})_{\bar{l}\nu_{4}}
   (P_{\bdq^{\prime}-\bdq})_{l\nu_{5}}(P_{\bdq^{\prime}-\bdq})_{\bar{l^{\prime}}\nu_{5}}\notag\\
  &+J_{3}(\bdq^{\prime}-\bdq)^{2}(P_{\bdq})_{\bar{l}\nu_{1}}(P_{\bdq})_{\bar{l^{\prime}}\nu_{2}}
   (P_{\bdq^{\prime}})_{\bar{l^{\prime}}\nu_{3}}(P_{\bdq^{\prime}})_{\bar{l}\nu_{4}}
   (P_{\bdq^{\prime}-\bdq})_{l\nu_{5}}(P_{\bdq^{\prime}-\bdq})_{l^{\prime}\nu_{5}}],\label{eq:V^a}\\
  \tilde{V}_{\nu_{1}\nu_{2}\nu_{3}\nu_{4}\nu_{5}}^{(b)}(\bdq,\bdq^{\prime})
  =-\frac{1}{4}\sum_{l,l^{\prime}=1}^{4}\textrm{sgn}(l)\textrm{sgn}(l^{\prime})
  &[J_{3}(\bdq^{\prime})^{2}(P_{\bdq})_{\bar{l^{\prime}}\nu_{1}}(P_{\bdq})_{\bar{l}\nu_{2}}
   (P_{\bdq^{\prime}})_{l\nu_{3}}(P_{\bdq^{\prime}})_{l^{\prime}\nu_{4}}  
   (P_{\bdq-\bdq^{\prime}})_{\bar{l}\nu_{5}}(P_{\bdq-\bdq^{\prime}})_{\bar{l^{\prime}}\nu_{5}}\notag\\[-11pt]
  &+J_{3}(\bdq^{\prime})J_{3}(\bdq-\bdq^{\prime})(P_{\bdq})_{\bar{l^{\prime}}\nu_{1}}(P_{\bdq})_{\bar{l}\nu_{2}}     
   (P_{\bdq^{\prime}})_{l\nu_{3}}(P_{\bdq^{\prime}})_{\bar{l^{\prime}}\nu_{4}}
   (P_{\bdq-\bdq^{\prime}})_{\bar{l}\nu_{5}}(P_{\bdq-\bdq^{\prime}})_{l^{\prime}\nu_{5}}\notag\\   
  &+J_{3}(\bdq-\bdq^{\prime})J_{3}(\bdq^{\prime})(P_{\bdq})_{\bar{l^{\prime}}\nu_{1}}(P_{\bdq})_{\bar{l}\nu_{2}}
   (P_{\bdq^{\prime}})_{\bar{l}\nu_{3}}(P_{\bdq^{\prime}})_{l^{\prime}\nu_{4}}
   (P_{\bdq-\bdq^{\prime}})_{l\nu_{5}}(P_{\bdq-\bdq^{\prime}})_{\bar{l^{\prime}}\nu_{5}}\notag\\  
  &+J_{3}(\bdq-\bdq^{\prime})^{2}(P_{\bdq})_{\bar{l^{\prime}}\nu_{1}}(P_{\bdq})_{\bar{l}\nu_{2}}    
   (P_{\bdq^{\prime}})_{\bar{l}\nu_{3}}(P_{\bdq^{\prime}})_{\bar{l^{\prime}}\nu_{4}}
   (P_{\bdq-\bdq^{\prime}})_{l\nu_{5}}(P_{\bdq-\bdq^{\prime}})_{l^{\prime}\nu_{5}}],\label{eq:V^b}\\
  \tilde{V}_{\nu_{1}\nu_{2}\nu_{3}\nu_{4}\nu_{5}}^{(c)}(\bdq,\bdq^{\prime})
  =-\frac{1}{4}\sum_{l,l^{\prime}=1}^{4}\textrm{sgn}(l)\textrm{sgn}(l^{\prime})
  &[J_{3}(\bdq)^{2}(P_{\bdq})_{l\nu_{1}}(P_{\bdq})_{l^{\prime}\nu_{2}}
    (P_{\bdq^{\prime}})_{\bar{l}\nu_{3}}(P_{\bdq^{\prime}})_{\bar{l^{\prime}}\nu_{4}}
    (P_{\bdq+\bdq^{\prime}})_{\bar{l^{\prime}}\nu_{5}}(P_{\bdq+\bdq^{\prime}})_{\bar{l}\nu_{5}}\notag\\[-11pt]
   &+J_{3}(\bdq)J_{3}(\bdq^{\prime})(P_{\bdq})_{l\nu_{1}}(P_{\bdq})_{\bar{l^{\prime}}\nu_{2}}
    (P_{\bdq^{\prime}})_{\bar{l}\nu_{3}}(P_{\bdq^{\prime}})_{l^{\prime}\nu_{4}}
    (P_{\bdq+\bdq^{\prime}})_{\bar{l^{\prime}}\nu_{5}}(P_{\bdq+\bdq^{\prime}})_{\bar{l}\nu_{5}}\notag\\
   &+J_{3}(\bdq^{\prime})J_{3}(\bdq)(P_{\bdq})_{\bar{l}\nu_{1}}(P_{\bdq})_{l^{\prime}\nu_{2}}
    (P_{\bdq^{\prime}})_{l\nu_{3}}(P_{\bdq^{\prime}})_{\bar{l^{\prime}}\nu_{4}}
    (P_{\bdq+\bdq^{\prime}})_{\bar{l^{\prime}}\nu_{5}}(P_{\bdq+\bdq^{\prime}})_{\bar{l}\nu_{5}}\notag\\
   &+J_{3}(\bdq^{\prime})^{2}(P_{\bdq})_{\bar{l}\nu_{1}}(P_{\bdq})_{\bar{l^{\prime}}\nu_{2}}
    (P_{\bdq^{\prime}})_{l\nu_{3}}(P_{\bdq^{\prime}})_{l^{\prime}\nu_{4}}
    (P_{\bdq+\bdq^{\prime}})_{\bar{l^{\prime}}\nu_{5}}(P_{\bdq+\bdq^{\prime}})_{\bar{l}\nu_{5}}],\label{eq:V^c}
\end{align}
and 
\begin{align}
  &f_{\nu_{1}\nu_{2}\nu_{3}\nu_{4}\nu_{5}}^{(a)}(\bdq,\bdq^{\prime};\tau,\tau_{1},\tau_{2})
  =G_{\nu_{1}}(\bdq,\tau_{1}-\tau)G_{\nu_{2}}(\bdq,\tau-\tau_{2})
  G_{\nu_{3}}(\bdq^{\prime},\tau_{2})G_{\nu_{4}}(\bdq^{\prime},-\tau_{1})
  G_{\nu_{5}}(\bdq^{\prime}-\bdq,\tau_{1}-\tau_{2}),\label{eq:f^a-tau}\\
  &f_{\nu_{1}\nu_{2}\nu_{3}\nu_{4}\nu_{5}}^{(b)}(\bdq,\bdq^{\prime};\tau,\tau_{1},\tau_{2})
  =G_{\nu_{1}}(\bdq,\tau_{2}-\tau)G_{\nu_{2}}(\bdq,\tau-\tau_{1})
  G_{\nu_{3}}(\bdq^{\prime},\tau_{1})G_{\nu_{4}}(\bdq^{\prime},-\tau_{2})
  G_{\nu_{5}}(\bdq-\bdq^{\prime},\tau_{1}-\tau_{2}),\label{eq:f^b-tau}\\
  &f_{\nu_{1}\nu_{2}\nu_{3}\nu_{4}\nu_{5}}^{(c)}(\bdq,\bdq^{\prime};\tau,\tau_{1},\tau_{2})
  =G_{\nu_{1}}(\bdq,\tau_{1}-\tau)G_{\nu_{2}}(\bdq,\tau-\tau_{2})
  G_{\nu_{3}}(\bdq^{\prime},\tau_{1})G_{\nu_{4}}(\bdq^{\prime},-\tau_{2})
  G_{\nu_{5}}(\bdq+\bdq^{\prime},\tau_{2}-\tau_{1}).\label{eq:f^c-tau}
\end{align}
By combining Eqs. (\ref{eq:DelG^II}){--}(\ref{eq:f^c-tau}) with Eq. (\ref{eq:Phi12-Ap})
and doing the integrals about $\tau$, $\tau_{1}$, and $\tau_{2}$ in Eq. (\ref{eq:DelG^II}),
we obtain
\begin{align}
  \Delta\Phi_{12}(i\Omega_{n})
  =\frac{1}{N}\sum_{\bdq,\bdq^{\prime}}\sum_{\nu_{1},\nu_{2},\nu_{3},\nu_{4},\nu_{5}=\alpha_{1},\beta_{1},\alpha_{2},\beta_{2}}
  v_{\nu_{1}\nu_{2}}^{z}(\bdq)e_{\nu_{3}\nu_{4}}^{z}(\bdq^{\prime})
  \sum_{k=a,b,c}
  \tilde{V}_{\nu_{1}\nu_{2}\nu_{3}\nu_{4}\nu_{5}}^{(k)}(\bdq,\bdq^{\prime})
  \tilde{I}_{\nu_{1}\nu_{2}\nu_{3}\nu_{4}\nu_{5}}^{(k)}(\bdq,\bdq^{\prime};i\Omega_{n}),\label{eq:DelPhi}
\end{align}
where
\begin{align}
  \tilde{I}_{\nu_{1}\nu_{2}\nu_{3}\nu_{4}\nu_{5}}^{(a)}(\bdq,\bdq^{\prime};i\Omega_{n})
  =T^{2}\sum_{m,m^{\prime}}G_{\nu_{1}}(\bdq,i\Omega_{m})G_{\nu_{2}}(\bdq,i\Omega_{n+m})
  G_{\nu_{3}}(\bdq^{\prime},i\Omega_{n+m^{\prime}})G_{\nu_{4}}(\bdq^{\prime},i\Omega_{m^{\prime}})
  G_{\nu_{5}}(\bdq^{\prime}-\bdq,i\Omega_{m^{\prime}-m}),\label{eq:tilI^a}\\
  \tilde{I}_{\nu_{1}\nu_{2}\nu_{3}\nu_{4}\nu_{5}}^{(b)}(\bdq,\bdq^{\prime};i\Omega_{n})
  =T^{2}\sum_{m,m^{\prime}}G_{\nu_{1}}(\bdq,i\Omega_{m})G_{\nu_{2}}(\bdq,i\Omega_{n+m})
  G_{\nu_{3}}(\bdq^{\prime},i\Omega_{n+m^{\prime}})G_{\nu_{4}}(\bdq^{\prime},i\Omega_{m^{\prime}})
  G_{\nu_{5}}(\bdq-\bdq^{\prime},i\Omega_{m-m^{\prime}}),\label{eq:tilI^b}\\
  \tilde{I}_{\nu_{1}\nu_{2}\nu_{3}\nu_{4}\nu_{5}}^{(c)}(\bdq,\bdq^{\prime};i\Omega_{n})
  =T^{2}\sum_{m,m^{\prime}}G_{\nu_{1}}(\bdq,i\Omega_{m})G_{\nu_{2}}(\bdq,i\Omega_{n+m})
  G_{\nu_{3}}(\bdq^{\prime},i\Omega_{m^{\prime}})G_{\nu_{4}}(\bdq^{\prime},i\Omega_{m^{\prime}-n})
  G_{\nu_{5}}(\bdq+\bdq^{\prime},i\Omega_{m+m^{\prime}}).\label{eq:tilI^c}
\end{align}
Then, to perform the analytic continuation,
we replace the Matsubara-frequency summations
in Eqs. (\ref{eq:tilI^a}){--}(\ref{eq:tilI^c})
by the corresponding integrals
in a way similar to that for metals~\cite{Eliashberg}. 
Namely,
since an intraband pair of the retarded and advanced Green's functions,
such as $G_{\nu}^{(\textrm{A})}(\bdq,z)G_{\nu}^{(\textrm{R})}(\bdq,z)$, gives
the leading contribution in the limit $\tau\rightarrow \infty$~\cite{Eliashberg},
we can express Eqs. (\ref{eq:tilI^a}){--}(\ref{eq:tilI^c})
in this limit as follows:
\begin{align}
  \tilde{I}_{\nu_{1}\nu_{2}\nu_{3}\nu_{4}\nu_{5}}^{(a)}(\bdq,\bdq^{\prime};i\Omega_{n})
  &\sim \delta_{\nu_{1},\nu_{2}}\delta_{\nu_{3},\nu_{4}}\int_{-\infty}^{\infty}\frac{dz}{2\pi i}n(z)
  [-G_{\nu_{1}}^{(\textrm{A})}(\bdq,z)
    G_{\nu_{2}}^{(\textrm{R})}(\bdq,z+i\Omega_{n})]\notag\\
  &\times 
  \int_{-\infty}^{\infty}\frac{dz^{\prime}}{2\pi i}n(z^{\prime})
  \{-G_{\nu_{3}}^{(\textrm{R})}(\bdq^{\prime},z^{\prime}+i\Omega_{n})
  G_{\nu_{4}}^{(\textrm{A})}(\bdq^{\prime},z^{\prime})
  G_{\nu_{5}}^{(\textrm{R})}(\bdq^{\prime}-\bdq,z^{\prime}-z)\notag\\
  &\ \ \ \ \ \ \ \ \ \ \ \ \ \ \ \ \ \ \ \ \
  +G_{\nu_{3}}^{(\textrm{R})}(\bdq^{\prime},z^{\prime}+z+i\Omega_{n})
  G_{\nu_{4}}^{(\textrm{A})}(\bdq^{\prime},z^{\prime}+z)
  [G_{\nu_{5}}^{(\textrm{R})}(\bdq^{\prime}-\bdq,z^{\prime})
    -G_{\nu_{5}}^{(\textrm{A})}(\bdq^{\prime}-\bdq,z^{\prime})]\notag\\
  &\ \ \ \ \ \ \ \ \ \ \ \ \ \ \ \ \ \ \ \ \
  +G_{\nu_{3}}^{(\textrm{R})}(\bdq^{\prime},z^{\prime})
  G_{\nu_{4}}^{(\textrm{A})}(\bdq^{\prime},z^{\prime}-i\Omega_{n})
  G_{\nu_{5}}^{(\textrm{A})}(\bdq^{\prime}-\bdq,z^{\prime}-z-i\Omega_{n})
  \}\notag\\
  &+\delta_{\nu_{1},\nu_{2}}\delta_{\nu_{3},\nu_{4}}\int_{-\infty}^{\infty}\frac{dz}{2\pi i}n(z)
  G_{\nu_{1}}^{(\textrm{A})}(\bdq,z-i\Omega_{n})
  G_{\nu_{2}}^{(\textrm{R})}(\bdq,z)\notag\\
  &\times 
  \int_{-\infty}^{\infty}\frac{dz^{\prime}}{2\pi i}n(z^{\prime})
  \{-G_{\nu_{3}}^{(\textrm{R})}(\bdq^{\prime},z^{\prime}+i\Omega_{n})
  G_{\nu_{4}}^{(\textrm{A})}(\bdq^{\prime},z^{\prime})
  G_{\nu_{5}}^{(\textrm{R})}(\bdq^{\prime}-\bdq,z^{\prime}-z+i\Omega_{n})\notag\\
  &\ \ \ \ \ \ \ \ \ \ \ \ \ \ \ \ \ \ \ \ \
  +G_{\nu_{3}}^{(\textrm{R})}(\bdq^{\prime},z^{\prime}+z)
  G_{\nu_{4}}^{(\textrm{A})}(\bdq^{\prime},z^{\prime}+z-i\Omega_{n})
  [G_{\nu_{5}}^{(\textrm{R})}(\bdq^{\prime}-\bdq,z^{\prime})
    -G_{\nu_{5}}^{(\textrm{A})}(\bdq^{\prime}-\bdq,z^{\prime})]\notag\\
  &\ \ \ \ \ \ \ \ \ \ \ \ \ \ \ \ \ \ \ \ \
  +G_{\nu_{3}}^{(\textrm{R})}(\bdq^{\prime},z^{\prime})
  G_{\nu_{4}}^{(\textrm{A})}(\bdq^{\prime},z^{\prime}-i\Omega_{n})
  G_{\nu_{5}}^{(\textrm{A})}(\bdq^{\prime}-\bdq,z^{\prime}-z)
  \},\label{eq:tilI^a-G}\\
  \tilde{I}_{\nu_{1}\nu_{2}\nu_{3}\nu_{4}\nu_{5}}^{(b)}(\bdq,\bdq^{\prime};i\Omega_{n})
  &\sim \delta_{\nu_{1},\nu_{2}}\delta_{\nu_{3},\nu_{4}}\int_{-\infty}^{\infty}\frac{dz}{2\pi i}n(z)
  [-G_{\nu_{1}}^{(\textrm{A})}(\bdq,z)
    G_{\nu_{2}}^{(\textrm{R})}(\bdq,z+i\Omega_{n})]\notag\\
  &\times 
  \int_{-\infty}^{\infty}\frac{dz^{\prime}}{2\pi i}n(z^{\prime})
  \{-G_{\nu_{3}}^{(\textrm{R})}(\bdq^{\prime},z^{\prime}+i\Omega_{n})
  G_{\nu_{4}}^{(\textrm{A})}(\bdq^{\prime},z^{\prime})
  G_{\nu_{5}}^{(\textrm{A})}(\bdq-\bdq^{\prime},z-z^{\prime})\notag\\
  &\ \ \ \ \ \ \ \ \ \ \ \ \ \ \ \ \ \ \ \ \
  -G_{\nu_{3}}^{(\textrm{R})}(\bdq^{\prime},z^{\prime}+z+i\Omega_{n})
  G_{\nu_{4}}^{(\textrm{A})}(\bdq^{\prime},z^{\prime}+z)
  [G_{\nu_{5}}^{(\textrm{R})}(\bdq-\bdq^{\prime},-z^{\prime})
    -G_{\nu_{5}}^{(\textrm{A})}(\bdq-\bdq^{\prime},-z^{\prime})]\notag\\
  &\ \ \ \ \ \ \ \ \ \ \ \ \ \ \ \ \ \ \ \ \
  +G_{\nu_{3}}^{(\textrm{R})}(\bdq^{\prime},z^{\prime})
  G_{\nu_{4}}^{(\textrm{A})}(\bdq^{\prime},z^{\prime}-i\Omega_{n})
  G_{\nu_{5}}^{(\textrm{R})}(\bdq-\bdq^{\prime},z-z^{\prime}+i\Omega_{n})
  \}\notag\\
  &+\delta_{\nu_{1},\nu_{2}}\delta_{\nu_{3},\nu_{4}}\int_{-\infty}^{\infty}\frac{dz}{2\pi i}n(z)
  G_{\nu_{1}}^{(\textrm{A})}(\bdq,z-i\Omega_{n})
  G_{\nu_{2}}^{(\textrm{R})}(\bdq,z)\notag\\
  &\times 
  \int_{-\infty}^{\infty}\frac{dz^{\prime}}{2\pi i}n(z^{\prime})
  \{-G_{\nu_{3}}^{(\textrm{R})}(\bdq^{\prime},z^{\prime}+i\Omega_{n})
  G_{\nu_{4}}^{(\textrm{A})}(\bdq^{\prime},z^{\prime})
  G_{\nu_{5}}^{(\textrm{A})}(\bdq-\bdq^{\prime},z-z^{\prime}-i\Omega_{n})\notag\\
  &\ \ \ \ \ \ \ \ \ \ \ \ \ \ \ \ \ \ \ \ \
  -G_{\nu_{3}}^{(\textrm{R})}(\bdq^{\prime},z^{\prime}+z)
  G_{\nu_{4}}^{(\textrm{A})}(\bdq^{\prime},z^{\prime}+z-i\Omega_{n})
  [G_{\nu_{5}}^{(\textrm{R})}(\bdq-\bdq^{\prime},-z^{\prime})
    -G_{\nu_{5}}^{(\textrm{A})}(\bdq-\bdq^{\prime},-z^{\prime})]\notag\\
  &\ \ \ \ \ \ \ \ \ \ \ \ \ \ \ \ \ \ \ \ \
  +G_{\nu_{3}}^{(\textrm{R})}(\bdq^{\prime},z^{\prime})
  G_{\nu_{4}}^{(\textrm{A})}(\bdq^{\prime},z^{\prime}-i\Omega_{n})
  G_{\nu_{5}}^{(\textrm{R})}(\bdq-\bdq^{\prime},z-z^{\prime})
  \},\label{eq:tilI^b-G}
\end{align}
\begin{align}
  \tilde{I}_{\nu_{1}\nu_{2}\nu_{3}\nu_{4}\nu_{5}}^{(c)}(\bdq,\bdq^{\prime};i\Omega_{n})
  &\sim \delta_{\nu_{1},\nu_{2}}\delta_{\nu_{3},\nu_{4}}\int_{-\infty}^{\infty}\frac{dz}{2\pi i}n(z)
  [-G_{\nu_{1}}^{(\textrm{A})}(\bdq,z)
    G_{\nu_{2}}^{(\textrm{R})}(\bdq,z+i\Omega_{n})]\notag\\
  &\times 
  \int_{-\infty}^{\infty}\frac{dz^{\prime}}{2\pi i}n(z^{\prime})
  \{-G_{\nu_{3}}^{(\textrm{R})}(\bdq^{\prime},z^{\prime}+i\Omega_{n})
  G_{\nu_{4}}^{(\textrm{A})}(\bdq^{\prime},z^{\prime})
  G_{\nu_{5}}^{(\textrm{R})}(\bdq^{\prime}+\bdq,z^{\prime}+z+i\Omega_{n})\notag\\
  &\ \ \ \ \ \ \ \ \ \ \ \ \ \ \ \ \ \ \ \ \
  +G_{\nu_{3}}^{(\textrm{R})}(\bdq^{\prime},z^{\prime}-z)
  G_{\nu_{4}}^{(\textrm{A})}(\bdq^{\prime},z^{\prime}-z-i\Omega_{n})
  [G_{\nu_{5}}^{(\textrm{R})}(\bdq^{\prime}+\bdq,z^{\prime})
    -G_{\nu_{5}}^{(\textrm{A})}(\bdq^{\prime}+\bdq,z^{\prime})]\notag\\
  &\ \ \ \ \ \ \ \ \ \ \ \ \ \ \ \ \ \ \ \ \
  +G_{\nu_{3}}^{(\textrm{R})}(\bdq^{\prime},z^{\prime})
  G_{\nu_{4}}^{(\textrm{A})}(\bdq^{\prime},z^{\prime}-i\Omega_{n})
  G_{\nu_{5}}^{(\textrm{A})}(\bdq^{\prime}+\bdq,z^{\prime}+z)
  \}\notag\\
  &+\delta_{\nu_{1},\nu_{2}}\delta_{\nu_{3},\nu_{4}}\int_{-\infty}^{\infty}\frac{dz}{2\pi i}n(z)
  G_{\nu_{1}}^{(\textrm{A})}(\bdq,z-i\Omega_{n})
  G_{\nu_{2}}^{(\textrm{R})}(\bdq,z)\notag\\
  &\times 
  \int_{-\infty}^{\infty}\frac{dz^{\prime}}{2\pi i}n(z^{\prime})
  \{-G_{\nu_{3}}^{(\textrm{R})}(\bdq^{\prime},z^{\prime}+i\Omega_{n})
  G_{\nu_{4}}^{(\textrm{A})}(\bdq^{\prime},z^{\prime})
  G_{\nu_{5}}^{(\textrm{R})}(\bdq^{\prime}+\bdq,z^{\prime}+z)\notag\\
  &\ \ \ \ \ \ \ \ \ \ \ \ \ \ \ \ \ \ \ \ \
  +G_{\nu_{3}}^{(\textrm{R})}(\bdq^{\prime},z^{\prime}-z+i\Omega_{n})
  G_{\nu_{4}}^{(\textrm{A})}(\bdq^{\prime},z^{\prime}-z)
  [G_{\nu_{5}}^{(\textrm{R})}(\bdq^{\prime}+\bdq,z^{\prime})
    -G_{\nu_{5}}^{(\textrm{A})}(\bdq^{\prime}+\bdq,z^{\prime})]\notag\\
  &\ \ \ \ \ \ \ \ \ \ \ \ \ \ \ \ \ \ \ \ \
  +G_{\nu_{3}}^{(\textrm{R})}(\bdq^{\prime},z^{\prime})
  G_{\nu_{4}}^{(\textrm{A})}(\bdq^{\prime},z^{\prime}-i\Omega_{n})
  G_{\nu_{5}}^{(\textrm{A})}(\bdq^{\prime}+\bdq,z^{\prime}+z-i\Omega_{n})
  \}.\label{eq:tilI^c-G}
\end{align}
In replacing the sums over $m$ in Eqs. (\ref{eq:tilI^a}){--}(\ref{eq:tilI^c})
by the contour integrals,
we have considered the contributions only from the region
for $-\Omega_{n}<\textrm{Im}z<0$ in the contour $C$ shown in Fig. \ref{fig4}(a)
because they include the pair of the retarded and advanced Green's functions.
Furthermore,
in replacing the sums over $m^{\prime}$ in Eqs. (\ref{eq:tilI^a}), (\ref{eq:tilI^b}),
and (\ref{eq:tilI^c}) by the integrals,
we have used the contours $C^{\prime}$, $C^{\prime}$, and $C^{\prime\prime}$,
respectively;
the $C^{\prime}$ and $C^{\prime\prime}$ are shown in Figs. \ref{fig4}(b) and \ref{fig4}(c).
We now perform the analytic continuation of Eqs. (\ref{eq:tilI^a-G}){--}(\ref{eq:tilI^c-G})
using the replacement $i\Omega_{n}\rightarrow\omega+i\delta$;
the results are 
\begin{align}
  \Delta I_{\nu\nu^{\prime}\nu^{\prime\prime}}^{\textrm{R}(a)}(\bdq,\bdq^{\prime};\omega)
  &=\tilde{I}_{\nu\nu\nu^{\prime}\nu^{\prime}\nu^{\prime\prime}}^{\textrm{R}(a)}(\bdq,\bdq^{\prime};\omega)
  -\tilde{I}_{\nu\nu\nu^{\prime}\nu^{\prime}\nu^{\prime\prime}}^{\textrm{R}(a)}(\bdq,\bdq^{\prime};0)\notag\\
  &\sim i\omega\int_{-\infty}^{\infty}\frac{dz}{2\pi}\frac{\partial n(z)}{\partial z}
  g_{\nu}(\bdq,z)
  \int_{-\infty}^{\infty}\frac{dz^{\prime}}{\pi}[n(z^{\prime})-n(z^{\prime}-z)]
  g_{\nu^{\prime}}(\bdq^{\prime},z^{\prime})
  \textrm{Im}G_{\nu^{\prime\prime}}^{(\textrm{R})}(\bdq^{\prime}-\bdq,z^{\prime}-z),\label{eq:Del-I^a}\\
  \Delta I_{\nu\nu^{\prime}\nu^{\prime\prime}}^{\textrm{R}(b)}(\bdq,\bdq^{\prime};\omega)
  &=\tilde{I}_{\nu\nu\nu^{\prime}\nu^{\prime}\nu^{\prime\prime}}^{\textrm{R}(b)}(\bdq,\bdq^{\prime};\omega)
  -\tilde{I}_{\nu\nu\nu^{\prime}\nu^{\prime}\nu^{\prime\prime}}^{\textrm{R}(b)}(\bdq,\bdq^{\prime};0)\notag\\
  &\sim i\omega\int_{-\infty}^{\infty}\frac{dz}{2\pi}\frac{\partial n(z)}{\partial z}
  g_{\nu}(\bdq,z)
  \int_{-\infty}^{\infty}\frac{dz^{\prime}}{\pi}[n(z^{\prime}-z)-n(z^{\prime})]
  g_{\nu^{\prime}}(\bdq^{\prime},z^{\prime})
  \textrm{Im}G_{\nu^{\prime\prime}}^{(\textrm{R})}(\bdq-\bdq^{\prime},z-z^{\prime}),\label{eq:Del-I^b}\\
  \Delta I_{\nu\nu^{\prime}\nu^{\prime\prime}}^{\textrm{R}(c)}(\bdq,\bdq^{\prime};\omega)
  &=\tilde{I}_{\nu\nu\nu^{\prime}\nu^{\prime}\nu^{\prime\prime}}^{\textrm{R}(c)}(\bdq,\bdq^{\prime};\omega)
  -\tilde{I}_{\nu\nu\nu^{\prime}\nu^{\prime}\nu^{\prime\prime}}^{\textrm{R}(c)}(\bdq,\bdq^{\prime};0)\notag\\
  &\sim i\omega\int_{-\infty}^{\infty}\frac{dz}{2\pi}\frac{\partial n(z)}{\partial z}
  g_{\nu}(\bdq,z)
  \int_{-\infty}^{\infty}\frac{dz^{\prime}}{\pi}[n(z^{\prime})-n(z^{\prime}+z)]
  g_{\nu^{\prime}}(\bdq^{\prime},z^{\prime})
  \textrm{Im}G_{\nu^{\prime\prime}}^{(\textrm{R})}(\bdq^{\prime}+\bdq,z^{\prime}+z),\label{eq:Del-I^c}
\end{align}
where we have introduced $g_{\nu}(\bdq,z)=G_{\nu}^{(\textrm{A})}(\bdq,z)G_{\nu}^{(\textrm{R})}(\bdq,z)$,
used $n(z)-n(z+\omega)\sim -\omega\frac{\partial n(z)}{\partial z}$,
and neglected the $O(\omega^{2})$ terms.
Combining Eqs. (\ref{eq:Del-I^a}){--}(\ref{eq:Del-I^c})
with Eq. (\ref{eq:DelPhi}) and
$\Delta\Phi_{12}^{\textrm{R}}(\omega)=\Delta\Phi_{12}(i\Omega_{n}\rightarrow \omega+i\delta)$,
we obtain
\begin{align}
  L_{12}^{\prime}=\lim_{\omega\rightarrow 0}
  \frac{\Delta\Phi^{\textrm{R}}_{12}(\omega)-\Delta\Phi^{\textrm{R}}_{12}(0)}{i\omega}
  =\frac{1}{N}\sum_{\bdq,\bdq^{\prime}}
  \sum_{\nu,\nu^{\prime},\nu^{\prime\prime}=\alpha_{1},\beta_{1},\alpha_{2},\beta_{2}}
  v_{\nu\nu}^{z}(\bdq)e_{\nu^{\prime}\nu^{\prime}}^{z}(\bdq^{\prime})
  \sum_{k=a,b,c}
  V_{\nu\nu^{\prime}\nu^{\prime\prime}}^{(k)}(\bdq,\bdq^{\prime})
  I_{\nu\nu^{\prime}\nu^{\prime\prime}}^{(k)}(\bdq,\bdq^{\prime}),\label{eq:L12'}
\end{align}
where
\begin{align}
  &V_{\nu\nu^{\prime}\nu^{\prime\prime}}^{(k)}(\bdq,\bdq^{\prime})
  =\tilde{V}_{\nu\nu\nu^{\prime}\nu^{\prime}\nu^{\prime\prime}}^{(k)}(\bdq,\bdq^{\prime}),\label{eq:V}\\
  &I_{\nu\nu^{\prime}\nu^{\prime\prime}}^{(a)}(\bdq,\bdq^{\prime})
  =\int_{-\infty}^{\infty}\frac{dz}{2\pi}\frac{\partial n(z)}{\partial z}
  g_{\nu}(\bdq,z)
  \int_{-\infty}^{\infty}\frac{dz^{\prime}}{\pi}[n(z^{\prime})-n(z^{\prime}-z)]
  g_{\nu^{\prime}}(\bdq^{\prime},z^{\prime})
  \textrm{Im}G_{\nu^{\prime\prime}}^{(\textrm{R})}(\bdq^{\prime}-\bdq,z^{\prime}-z),\label{eq:I^a}\\
  &I_{\nu\nu^{\prime}\nu^{\prime\prime}}^{(b)}(\bdq,\bdq^{\prime})
  =\int_{-\infty}^{\infty}\frac{dz}{2\pi}\frac{\partial n(z)}{\partial z}
  g_{\nu}(\bdq,z)
  \int_{-\infty}^{\infty}\frac{dz^{\prime}}{\pi}[n(z^{\prime}-z)-n(z^{\prime})]
  g_{\nu^{\prime}}(\bdq^{\prime},z^{\prime})
  \textrm{Im}G_{\nu^{\prime\prime}}^{(\textrm{R})}(\bdq-\bdq^{\prime},z-z^{\prime}),\label{eq:I^b}\\
  &I_{\nu\nu^{\prime}\nu^{\prime\prime}}^{(c)}(\bdq,\bdq^{\prime})
  =\int_{-\infty}^{\infty}\frac{dz}{2\pi}\frac{\partial n(z)}{\partial z}
  g_{\nu}(\bdq,z)
  \int_{-\infty}^{\infty}\frac{dz^{\prime}}{\pi}[n(z^{\prime})-n(z^{\prime}+z)]
  g_{\nu^{\prime}}(\bdq^{\prime},z^{\prime})
  \textrm{Im}G_{\nu^{\prime\prime}}^{(\textrm{R})}(\bdq^{\prime}+\bdq,z^{\prime}+z).\label{eq:I^c}
\end{align}
Note that
$\tilde{V}_{\nu\nu\nu^{\prime}\nu^{\prime}\nu^{\prime\prime}}^{(k)}(\bdq,\bdq^{\prime})$'s
have been given by Eqs. (\ref{eq:V^a}){--}(\ref{eq:V^c}). 
In the limit $\tau=1/2\gamma\rightarrow \infty$,
we can easily do the integrals in Eqs. (\ref{eq:I^a}){--}(\ref{eq:I^c})
by using the approximate relations,
\begin{align}
  &g_{\nu}(\bdq,z)=G_{\nu}^{(\textrm{A})}(\bdq,z)G_{\nu}^{(\textrm{R})}(\bdq,z)
  =\frac{1}{[z+(-1)^{\nu}\epsilon_{\nu}(\bdq)]^{2}+\gamma^{2}}
  \sim \frac{\pi}{\gamma}\delta[z+(-1)^{\nu}\epsilon_{\nu}(\bdq)],\label{eq:g-approx}\\
  &\textrm{Im}G_{\nu}^{(\textrm{R})}(\bdq,z)
  =(-1)^{\nu}\frac{\gamma}{[z+(-1)^{\nu}\epsilon_{\nu}(\bdq)]^{2}+\gamma^{2}}
  \sim (-1)^{\nu}\pi\delta[z+(-1)^{\nu}\epsilon_{\nu}(\bdq)],\label{eq:ImG-approx}
\end{align}
where $(-1)^{\nu}=-1$ for $\nu=\alpha_{1},\beta_{1}$ and $1$ for $\nu=\alpha_{2},\beta_{2}$.
Combining these equations with Eqs. (\ref{eq:I^a}){--}(\ref{eq:I^c}),
we obtain
\begin{align}
  I_{\nu\nu^{\prime}\nu^{\prime\prime}}^{(a)}(\bdq,\bdq^{\prime})
  \sim& \frac{\pi}{2\gamma^{2}}
  \frac{\partial n[\epsilon_{\nu}(\bdq)]}{\partial \epsilon_{\nu}(\bdq)}
  \{n[(-1)^{\nu^{\prime}+1}\epsilon_{\nu^{\prime}}(\bdq^{\prime})]
  -n[(-1)^{\nu^{\prime\prime}+1}\epsilon_{\nu^{\prime\prime}}(\bdq^{\prime}-\bdq)]
  \}
  (-1)^{\nu^{\prime\prime}}\notag\\
  &\times 
  \delta[(-1)^{\nu}\epsilon_{\nu}(\bdq)-(-1)^{\nu^{\prime}}\epsilon_{\nu^{\prime}}(\bdq^{\prime})
    +(-1)^{\nu^{\prime\prime}}\epsilon_{\nu^{\prime\prime}}(\bdq^{\prime}-\bdq)],\label{eq:I^a-approx}\\
  I_{\nu\nu^{\prime}\nu^{\prime\prime}}^{(b)}(\bdq,\bdq^{\prime})
  \sim& \frac{\pi}{2\gamma^{2}}
  \frac{\partial n[\epsilon_{\nu}(\bdq)]}{\partial \epsilon_{\nu}(\bdq)}
  \{n[(-1)^{\nu^{\prime\prime}}\epsilon_{\nu^{\prime\prime}}(\bdq-\bdq^{\prime})]
  -n[(-1)^{\nu^{\prime}+1}\epsilon_{\nu^{\prime}}(\bdq^{\prime})]
  \}
  (-1)^{\nu^{\prime\prime}}\notag\\
  &\times 
  \delta[(-1)^{\nu}\epsilon_{\nu}(\bdq)-(-1)^{\nu^{\prime}}\epsilon_{\nu^{\prime}}(\bdq^{\prime})
    -(-1)^{\nu^{\prime\prime}}\epsilon_{\nu^{\prime\prime}}(\bdq-\bdq^{\prime})],\label{eq:I^b-approx}\\
  I_{\nu\nu^{\prime}\nu^{\prime\prime}}^{(c)}(\bdq,\bdq^{\prime})
  \sim& \frac{\pi}{2\gamma^{2}}
  \frac{\partial n[\epsilon_{\nu}(\bdq)]}{\partial \epsilon_{\nu}(\bdq)}
  \{n[(-1)^{\nu^{\prime}+1}\epsilon_{\nu^{\prime}}(\bdq^{\prime})]
  -n[(-1)^{\nu^{\prime\prime}+1}\epsilon_{\nu^{\prime\prime}}(\bdq^{\prime}+\bdq)]
  \}
  (-1)^{\nu^{\prime\prime}}\notag\\
  &\times 
  \delta[(-1)^{\nu}\epsilon_{\nu}(\bdq)+(-1)^{\nu^{\prime}}\epsilon_{\nu^{\prime}}(\bdq^{\prime})
    -(-1)^{\nu^{\prime\prime}}\epsilon_{\nu^{\prime\prime}}(\bdq^{\prime}+\bdq)],\label{eq:I^c-approx}
\end{align}
where the delta functions represent the energy conservation relations 
in the scattering processes due to the second-order $H_{\textrm{int}}$. 
These equations can be obtained also 
by using Eqs. (\ref{eq:G^R-1}) and (\ref{eq:G^R-2})
and the relation $\frac{3}{x^{2}+(3\gamma)^{2}}\sim \frac{\pi}{\gamma}\delta(x)$,
instead of Eqs. (\ref{eq:g-approx}) and (\ref{eq:ImG-approx}),
and doing the integrals in Eqs. (\ref{eq:I^a}){--}(\ref{eq:I^c}).
This is the reason why we have used that relation about the Lorentzian function
in the numerical evaluations of
$S_{\textrm{m}}$, $\sigma_{\textrm{m}}$, and $\kappa_{\textrm{m}}$.
Then,
performing some calculations using Eqs. (\ref{eq:V}), (\ref{eq:V^a}){--}(\ref{eq:V^c}), and
(\ref{eq:P-1}){--}(\ref{eq:P-4}),
we find that the finite terms of $V_{\nu\nu^{\prime}\nu^{\prime\prime}}^{(p)}(\bdq,\bdq^{\prime})$'s ($p=1,2,3$)
are given by those for $(\nu,\nu^{\prime},\nu^{\prime\prime})=(\beta,\beta,\beta)$,
$(\beta,\alpha,\alpha)$, $(\alpha,\beta,\alpha)$, and $(\alpha,\alpha,\beta)$,
which are expressed as follows:
\begin{align}
  &V_{\nu\nu^{\prime}\nu^{\prime\prime}}^{(1)}(\bdq,\bdq^{\prime})
  =V_{\nu_{1}\nu^{\prime}_{1}\nu^{\prime\prime}_{2}}^{(a)}(\bdq,\bdq^{\prime})
  +V_{\nu_{2}\nu^{\prime}_{2}\nu^{\prime\prime}_{1}}^{(a)}(\bdq,\bdq^{\prime})
  +V_{\nu_{1}\nu^{\prime}_{1}\nu^{\prime\prime}_{1}}^{(b)}(\bdq,\bdq^{\prime})
  +V_{\nu_{2}\nu^{\prime}_{2}\nu^{\prime\prime}_{2}}^{(b)}(\bdq,\bdq^{\prime})
  +V_{\nu_{1}\nu^{\prime}_{2}\nu^{\prime\prime}_{1}}^{(c)}(\bdq,-\bdq^{\prime})
  +V_{\nu_{2}\nu^{\prime}_{1}\nu^{\prime\prime}_{2}}^{(c)}(\bdq,-\bdq^{\prime}),\label{eq:V_1}\\
  &V_{\nu\nu^{\prime}\nu^{\prime\prime}}^{(2)}(\bdq,\bdq^{\prime})
  =V_{\nu_{1}\nu^{\prime}_{1}\nu^{\prime\prime}_{1}}^{(a)}(\bdq,\bdq^{\prime})
  +V_{\nu_{2}\nu^{\prime}_{2}\nu^{\prime\prime}_{2}}^{(a)}(\bdq,\bdq^{\prime})
  +V_{\nu_{1}\nu^{\prime}_{1}\nu^{\prime\prime}_{2}}^{(b)}(\bdq,\bdq^{\prime})
  +V_{\nu_{2}\nu^{\prime}_{2}\nu^{\prime\prime}_{1}}^{(b)}(\bdq,\bdq^{\prime})
  +V_{\nu_{2}\nu^{\prime}_{1}\nu^{\prime\prime}_{1}}^{(c)}(\bdq,-\bdq^{\prime})
  +V_{\nu_{1}\nu^{\prime}_{2}\nu^{\prime\prime}_{2}}^{(c)}(\bdq,-\bdq^{\prime}),\label{eq:V_2}\\
  &V_{\nu\nu^{\prime}\nu^{\prime\prime}}^{(3)}(\bdq,\bdq^{\prime})
  =V_{\nu_{2}\nu^{\prime}_{1}\nu^{\prime\prime}_{1}}^{(a)}(\bdq,\bdq^{\prime})
  +V_{\nu_{1}\nu^{\prime}_{2}\nu^{\prime\prime}_{2}}^{(a)}(\bdq,\bdq^{\prime})
  +V_{\nu_{1}\nu^{\prime}_{2}\nu^{\prime\prime}_{1}}^{(b)}(\bdq,\bdq^{\prime})
  +V_{\nu_{2}\nu^{\prime}_{1}\nu^{\prime\prime}_{2}}^{(b)}(\bdq,\bdq^{\prime})
  +V_{\nu_{1}\nu^{\prime}_{1}\nu^{\prime\prime}_{1}}^{(c)}(\bdq,-\bdq^{\prime})
  +V_{\nu_{2}\nu^{\prime}_{2}\nu^{\prime\prime}_{2}}^{(c)}(\bdq,-\bdq^{\prime}).\label{eq:V_3}
\end{align}
[Note that if $(\nu,\nu^{\prime},\nu^{\prime\prime})=(\beta,\alpha,\alpha)$,
we have $(\nu_{1},\nu^{\prime}_{1},\nu^{\prime\prime}_{2})=(\beta_{1},\alpha_{1},\alpha_{2})$,
$(\nu_{2},\nu^{\prime}_{2},\nu^{\prime\prime}_{1})=(\beta_{2},\alpha_{2},\alpha_{1})$,
etc.] 
Since
$V_{\nu\nu^{\prime}\nu^{\prime\prime}}^{(k)}(\bdq,\bdq^{\prime})$'s ($k=a,b,c$)
include the square of the coupling constant of $H_{\textrm{int}}$
[see Eqs. (\ref{eq:V^a}){--}(\ref{eq:V^c}) with Eq. (\ref{eq:V})]
and $J_{3}(\bdq)=\sqrt{\frac{4S}{N}}\sin2\phi J(\bdq)$, 
we can write the finite terms of $V_{\nu\nu^{\prime}\nu^{\prime\prime}}^{(p)}(\bdq,\bdq^{\prime})$'s
($p=1,2,3$) as follows:
\begin{align}
  V_{\nu\nu^{\prime}\nu^{\prime\prime}}^{(p)}(\bdq,\bdq^{\prime})
  =v_{\nu\nu^{\prime}\nu^{\prime\prime}}^{(p)}(\bdq,\bdq^{\prime})\frac{S}{2N}\sin^{2}2\phi,\label{eq:V-v}
\end{align}
where
\begin{align}
  &v_{\beta\beta\beta}^{(1)}(\bdq,\bdq^{\prime})
  =+v_{a0}(\bdq,\bdq^{\prime})C_{\bdq}^{\prime}
  -v_{b0}(\bdq,\bdq^{\prime})C_{\bdq^{\prime}}^{\prime}
  -v_{c0}(\bdq,\bdq^{\prime})C_{\bdq-\bdq^{\prime}}^{\prime}
  -v_{d0}(\bdq,\bdq^{\prime})
  (C_{\bdq}^{\prime}C_{\bdq^{\prime}}^{\prime}C_{\bdq-\bdq^{\prime}}^{\prime}
  +S_{\bdq}^{\prime}S_{\bdq^{\prime}}^{\prime}S_{\bdq-\bdq^{\prime}}^{\prime}),
  \label{eq:vbbb^1}\\
  &v_{\beta\beta\beta}^{(2)}(\bdq,\bdq^{\prime})
  =-v_{a0}(\bdq,\bdq^{\prime})C_{\bdq}^{\prime}
  +v_{b0}(\bdq,\bdq^{\prime})C_{\bdq^{\prime}}^{\prime}
  -v_{c0}(\bdq,\bdq^{\prime})C_{\bdq-\bdq^{\prime}}^{\prime}
  -v_{d0}(\bdq,\bdq^{\prime})
  (C_{\bdq}^{\prime}C_{\bdq^{\prime}}^{\prime}C_{\bdq-\bdq^{\prime}}^{\prime}
  +S_{\bdq}^{\prime}S_{\bdq^{\prime}}^{\prime}S_{\bdq-\bdq^{\prime}}^{\prime})
  \label{eq:vbbb^2},\\
  &v_{\beta\beta\beta}^{(3)}(\bdq,\bdq^{\prime})
  =-v_{a0}(\bdq,\bdq^{\prime})C_{\bdq}^{\prime}
  -v_{b0}(\bdq,\bdq^{\prime})C_{\bdq^{\prime}}^{\prime}
  +v_{c0}(\bdq,\bdq^{\prime})C_{\bdq-\bdq^{\prime}}^{\prime}
  -v_{d0}(\bdq,\bdq^{\prime})
  (C_{\bdq}^{\prime}C_{\bdq^{\prime}}^{\prime}C_{\bdq-\bdq^{\prime}}^{\prime}
  +S_{\bdq}^{\prime}S_{\bdq^{\prime}}^{\prime}S_{\bdq-\bdq^{\prime}}^{\prime})
  \label{eq:vbbb^3},\\
  &v_{\beta\alpha\alpha}^{(1)}(\bdq,\bdq^{\prime})
  =+v_{a1}(\bdq,\bdq^{\prime})C_{\bdq}^{\prime}
  -v_{b1}(\bdq,\bdq^{\prime})C_{\bdq^{\prime}}
  -v_{c1}(\bdq,\bdq^{\prime})C_{\bdq-\bdq^{\prime}}
  -v_{d1}(\bdq,\bdq^{\prime})
  (C_{\bdq}^{\prime}C_{\bdq^{\prime}}C_{\bdq-\bdq^{\prime}}
  +S_{\bdq}^{\prime}S_{\bdq^{\prime}}S_{\bdq-\bdq^{\prime}}),
  \label{eq:vbaa^1}\\
  &v_{\beta\alpha\alpha}^{(2)}(\bdq,\bdq^{\prime})
  =-v_{a1}(\bdq,\bdq^{\prime})C_{\bdq}^{\prime}
  +v_{b1}(\bdq,\bdq^{\prime})C_{\bdq^{\prime}}
  -v_{c1}(\bdq,\bdq^{\prime})C_{\bdq-\bdq^{\prime}}
  -v_{d1}(\bdq,\bdq^{\prime})
  (C_{\bdq}^{\prime}C_{\bdq^{\prime}}C_{\bdq-\bdq^{\prime}}
  +S_{\bdq}^{\prime}S_{\bdq^{\prime}}S_{\bdq-\bdq^{\prime}}),
  \label{eq:vbaa^2}\\
  &v_{\beta\alpha\alpha}^{(3)}(\bdq,\bdq^{\prime})
  =-v_{a1}(\bdq,\bdq^{\prime})C_{\bdq}^{\prime}
  -v_{b1}(\bdq,\bdq^{\prime})C_{\bdq^{\prime}}
  +v_{c1}(\bdq,\bdq^{\prime})C_{\bdq-\bdq^{\prime}}
  -v_{d1}(\bdq,\bdq^{\prime})
  (C_{\bdq}^{\prime}C_{\bdq^{\prime}}C_{\bdq-\bdq^{\prime}}
  +S_{\bdq}^{\prime}S_{\bdq^{\prime}}S_{\bdq-\bdq^{\prime}}),
  \label{eq:vbaa^3}\\
  &v_{\alpha\beta\alpha}^{(1)}(\bdq,\bdq^{\prime})
  =+v_{a2}(\bdq,\bdq^{\prime})C_{\bdq}
  -v_{b2}(\bdq,\bdq^{\prime})C_{\bdq^{\prime}}^{\prime}
  -v_{c2}(\bdq,\bdq^{\prime})C_{\bdq-\bdq^{\prime}}
  -v_{d2}(\bdq,\bdq^{\prime})
  (C_{\bdq}C_{\bdq^{\prime}}^{\prime}C_{\bdq-\bdq^{\prime}}
  +S_{\bdq}S_{\bdq^{\prime}}^{\prime}S_{\bdq-\bdq^{\prime}}),
  \label{eq:vaba^1}\\
  &v_{\alpha\beta\alpha}^{(2)}(\bdq,\bdq^{\prime})
  =-v_{a2}(\bdq,\bdq^{\prime})C_{\bdq}
  +v_{b2}(\bdq,\bdq^{\prime})C_{\bdq^{\prime}}^{\prime}
  -v_{c2}(\bdq,\bdq^{\prime})C_{\bdq-\bdq^{\prime}}
  -v_{d2}(\bdq,\bdq^{\prime})
  (C_{\bdq}C_{\bdq^{\prime}}^{\prime}C_{\bdq-\bdq^{\prime}}
  +S_{\bdq}S_{\bdq^{\prime}}^{\prime}S_{\bdq-\bdq^{\prime}}),
  \label{eq:vaba^2}\\
  &v_{\alpha\beta\alpha}^{(3)}(\bdq,\bdq^{\prime})
  =-v_{a2}(\bdq,\bdq^{\prime})C_{\bdq}
  -v_{b2}(\bdq,\bdq^{\prime})C_{\bdq^{\prime}}^{\prime}
  +v_{c2}(\bdq,\bdq^{\prime})C_{\bdq-\bdq^{\prime}}
  -v_{d2}(\bdq,\bdq^{\prime})
  (C_{\bdq}C_{\bdq^{\prime}}^{\prime}C_{\bdq-\bdq^{\prime}}
  +S_{\bdq}S_{\bdq^{\prime}}^{\prime}S_{\bdq-\bdq^{\prime}}),
  \label{eq:vaba^3}\\
  &v_{\alpha\alpha\beta}^{(1)}(\bdq,\bdq^{\prime})
  =+v_{a3}(\bdq,\bdq^{\prime})C_{\bdq}
  -v_{b3}(\bdq,\bdq^{\prime})C_{\bdq^{\prime}}
  -v_{c3}(\bdq,\bdq^{\prime})C_{\bdq-\bdq^{\prime}}^{\prime}
  -v_{d3}(\bdq,\bdq^{\prime})
  (C_{\bdq}C_{\bdq^{\prime}}C_{\bdq-\bdq^{\prime}}^{\prime}
  +S_{\bdq}S_{\bdq^{\prime}}S_{\bdq-\bdq^{\prime}}^{\prime}),
  \label{eq:vaab^1}\\
  &v_{\alpha\alpha\beta}^{(2)}(\bdq,\bdq^{\prime})
  =-v_{a3}(\bdq,\bdq^{\prime})C_{\bdq}
  +v_{b3}(\bdq,\bdq^{\prime})C_{\bdq^{\prime}}
  -v_{c3}(\bdq,\bdq^{\prime})C_{\bdq-\bdq^{\prime}}^{\prime}
  -v_{d3}(\bdq,\bdq^{\prime})
  (C_{\bdq}C_{\bdq^{\prime}}C_{\bdq-\bdq^{\prime}}^{\prime}
  +S_{\bdq}S_{\bdq^{\prime}}S_{\bdq-\bdq^{\prime}}^{\prime}),
  \label{eq:vaab^2}\\
  &v_{\alpha\alpha\beta}^{(3)}(\bdq,\bdq^{\prime})
  =-v_{a3}(\bdq,\bdq^{\prime})C_{\bdq}
  -v_{b3}(\bdq,\bdq^{\prime})C_{\bdq^{\prime}}
  +v_{c3}(\bdq,\bdq^{\prime})C_{\bdq-\bdq^{\prime}}^{\prime}
  -v_{d3}(\bdq,\bdq^{\prime})
  (C_{\bdq}C_{\bdq^{\prime}}C_{\bdq-\bdq^{\prime}}^{\prime}
  +S_{\bdq}S_{\bdq^{\prime}}S_{\bdq-\bdq^{\prime}}^{\prime}),
  \label{eq:vaab^3}
\end{align}
and 
\begin{align}
  &C_{\bdq}^{\prime}=\cosh2\theta_{\bdq}^{\prime},\
  S_{\bdq}^{\prime}=\sinh2\theta_{\bdq}^{\prime},\
  C_{\bdq}=\cosh2\theta_{\bdq},\ 
  S_{\bdq}=\sinh2\theta_{\bdq},\label{eq:C-S}\\
  &v_{a0}(\bdq,\bdq^{\prime})
  =J(\bdq)[J(\bdq)+J(\bdq^{\prime})]
  +J(\bdq-\bdq^{\prime})[J(\bdq)-J(\bdq^{\prime})],\label{eq:v^a0}\\
  &v_{b0}(\bdq,\bdq^{\prime})
  =J(\bdq^{\prime})[J(\bdq^{\prime})+J(\bdq)]
  -J(\bdq-\bdq^{\prime})[J(\bdq)-J(\bdq^{\prime})],\label{eq:v^b0}\\
  &v_{c0}(\bdq,\bdq^{\prime})
  =J(\bdq-\bdq^{\prime})[J(\bdq)+J(\bdq^{\prime})+J(\bdq-\bdq^{\prime})]
  -J(\bdq)J(\bdq^{\prime}),\label{eq:v^c0}\\
  &v_{d0}(\bdq,\bdq^{\prime})
  =[J(\bdq)+J(\bdq^{\prime})]^{2}-J(\bdq)J(\bdq^{\prime})
  +J(\bdq-\bdq^{\prime})[J(\bdq)+J(\bdq^{\prime})+J(\bdq-\bdq^{\prime})],\label{eq:v^d0}\\
  &v_{a1}(\bdq,\bdq^{\prime})
  =J(\bdq)[J(\bdq)-J(\bdq^{\prime})]
  -J(\bdq-\bdq^{\prime})[J(\bdq)+J(\bdq^{\prime})],\label{eq:v^a1}\\
  &v_{b1}(\bdq,\bdq^{\prime})
  =J(\bdq^{\prime})[J(\bdq^{\prime})-J(\bdq)]
  +J(\bdq-\bdq^{\prime})[J(\bdq)+J(\bdq^{\prime})],\label{eq:v^b1}\\
  &v_{c1}(\bdq,\bdq^{\prime})
  =J(\bdq-\bdq^{\prime})[J(\bdq-\bdq^{\prime})-J(\bdq)+J(\bdq^{\prime})]
  +J(\bdq)J(\bdq^{\prime}),\label{eq:v^c1}\\
  &v_{d1}(\bdq,\bdq^{\prime})
  =[J(\bdq)-J(\bdq^{\prime})]^{2}+J(\bdq)J(\bdq^{\prime})
  +J(\bdq-\bdq^{\prime})[J(\bdq-\bdq^{\prime})-J(\bdq)+J(\bdq^{\prime})],\label{eq:v^d1}\\
  &v_{a2}(\bdq,\bdq^{\prime})
  =J(\bdq)[J(\bdq)-J(\bdq^{\prime})]
  +J(\bdq-\bdq^{\prime})[J(\bdq)+J(\bdq^{\prime})],\label{eq:v^a2}\\
  &v_{b2}(\bdq,\bdq^{\prime})
  =J(\bdq^{\prime})[J(\bdq^{\prime})-J(\bdq)]
  -J(\bdq-\bdq^{\prime})[J(\bdq)+J(\bdq^{\prime})],\label{eq:v^b2}\\
  &v_{c2}(\bdq,\bdq^{\prime})
  =J(\bdq-\bdq^{\prime})[J(\bdq)-J(\bdq^{\prime})+J(\bdq-\bdq^{\prime})]
  +J(\bdq)J(\bdq^{\prime}),\label{eq:v^c2}\\
  &v_{d2}(\bdq,\bdq^{\prime})
  =[J(\bdq)-J(\bdq^{\prime})]^{2}+J(\bdq)J(\bdq^{\prime})
  +J(\bdq-\bdq^{\prime})[J(\bdq)-J(\bdq^{\prime})+J(\bdq-\bdq^{\prime})],\label{eq:v^d2}\\
  &v_{a3}(\bdq,\bdq^{\prime})
  =J(\bdq)[J(\bdq)+J(\bdq^{\prime})]
  -J(\bdq-\bdq^{\prime})[J(\bdq)-J(\bdq^{\prime})],\label{eq:v^a3}\\
  &v_{b3}(\bdq,\bdq^{\prime})
  =J(\bdq^{\prime})[J(\bdq^{\prime})+J(\bdq)]
  +J(\bdq-\bdq^{\prime})[J(\bdq)-J(\bdq^{\prime})],\label{eq:v^b3}\\
  &v_{c3}(\bdq,\bdq^{\prime})
  =J(\bdq-\bdq^{\prime})[J(\bdq-\bdq^{\prime})-J(\bdq)-J(\bdq^{\prime})]
  -J(\bdq)J(\bdq^{\prime}),\label{eq:v^c3}\\
  &v_{d3}(\bdq,\bdq^{\prime})
  =[J(\bdq)+J(\bdq^{\prime})]^{2}-J(\bdq)J(\bdq^{\prime})
  -J(\bdq-\bdq^{\prime})[J(\bdq)+J(\bdq^{\prime})-J(\bdq-\bdq^{\prime})].\label{eq:v^d3}
\end{align}
[Note that the hyperbolic functions Eq. (\ref{eq:C-S}) satisfy
$\tanh2\theta_{\bdq}=-\frac{B^{\prime}(\bdq)}{A+A^{\prime}(\bdq)}$
and $\tanh2\theta^{\prime}_{\bdq}=\frac{B^{\prime}(\bdq)}{A-A^{\prime}(\bdq)}$,
as described in Sec. II B.] 
Equations (\ref{eq:vbbb^1}){--}(\ref{eq:vaab^3})
with Eqs. (\ref{eq:C-S}){--}(\ref{eq:v^d3}) give
the expressions of the $v^{(p)}_{\nu\nu^{\prime}\nu^{\prime\prime}}(\bdq,\bdq^{\prime})$'s
($p=1,2,3$) appearing in Eqs. (\ref{eq:F^2}){--}(\ref{eq:F^1}).
By combining Eqs. (\ref{eq:V-v}){--}(\ref{eq:v^d3}),
(\ref{eq:I^a-approx}){--}(\ref{eq:I^c-approx}),
and (\ref{eq:v-band}){--}(\ref{eq:e-band2})
with Eq. (\ref{eq:L12'}), 
we can express $L_{12}^{\prime}$ in the limit $\tau\rightarrow \infty$ as follows:
\begin{align}
  L_{12}^{\prime}=&\frac{\pi}{N^{2}}\sum_{\bdq,\bdq^{\prime}}
  \sum_{\nu,\nu^{\prime},\nu^{\prime\prime}=\alpha,\beta}
  v_{\nu\nu}^{z}(\bdq)e_{\nu^{\prime}\nu^{\prime}}^{z}(\bdq^{\prime})\tau^{2}
  \frac{\partial n[\epsilon_{\nu}(\bdq)]}{\partial \epsilon_{\nu}(\bdq)}
  S\sin^{2}2\phi \sum_{p=1,2,3}
  F^{(p)}_{\nu\nu^{\prime}\nu^{\prime\prime}}(\bdq,\bdq^{\prime}),\label{eq:L12'-last}
\end{align}
where
\begin{align}
  &F_{\nu\nu^{\prime}\nu^{\prime\prime}}^{(1)}(\bdq,\bdq^{\prime})
  =v_{\nu\nu^{\prime}\nu^{\prime\prime}}^{(1)}(\bdq,\bdq^{\prime})
  \{1+n[\epsilon_{\nu^{\prime\prime}}(\bdq-\bdq^{\prime})]+n[\epsilon_{\nu^{\prime}}(\bdq^{\prime})]\}
  \delta[\epsilon_{\nu}(\bdq)-\epsilon_{\nu^{\prime}}(\bdq^{\prime})
    -\epsilon_{\nu^{\prime\prime}}(\bdq-\bdq^{\prime})],\label{eq:F^1-Ap}\\ 
  &F_{\nu\nu^{\prime}\nu^{\prime\prime}}^{(2)}(\bdq,\bdq^{\prime})
  =v_{\nu\nu^{\prime}\nu^{\prime\prime}}^{(2)}(\bdq,\bdq^{\prime})
  \{n[\epsilon_{\nu^{\prime\prime}}(\bdq-\bdq^{\prime})]-n[\epsilon_{\nu^{\prime}}(\bdq^{\prime})]\}
  \delta[\epsilon_{\nu}(\bdq)-\epsilon_{\nu^{\prime}}(\bdq^{\prime})
    +\epsilon_{\nu^{\prime\prime}}(\bdq-\bdq^{\prime})],\label{eq:F^2-Ap}\\
  &F_{\nu\nu^{\prime}\nu^{\prime\prime}}^{(3)}(\bdq,\bdq^{\prime})
  =-v_{\nu\nu^{\prime}\nu^{\prime\prime}}^{(3)}(\bdq,\bdq^{\prime})
  \{n[\epsilon_{\nu^{\prime\prime}}(\bdq-\bdq^{\prime})]-n[\epsilon_{\nu^{\prime}}(\bdq^{\prime})]\}
  \delta[\epsilon_{\nu}(\bdq)+\epsilon_{\nu^{\prime}}(\bdq^{\prime})
    -\epsilon_{\nu^{\prime\prime}}(\bdq-\bdq^{\prime})].\label{eq:F^3-Ap}
\end{align}
In deriving them,
we have used the identity $n(-x)=-1-n(x)$.
Then, since Eqs. (\ref{eq:JS,JE}) and (\ref{eq:Phi12}){--}(\ref{eq:Phi22})
show that
$L_{11}^{\prime}$ and $L_{22}^{\prime}$ are obtained
by replacing $e_{\nu^{\prime}\nu^{\prime}}^{z}(\bdq^{\prime})$ in Eq. (\ref{eq:L12'-last})
by $v_{\nu^{\prime}\nu^{\prime}}^{z}(\bdq^{\prime})$
and by replacing $v_{\nu\nu}^{z}(\bdq)$ in Eq. (\ref{eq:L12'-last})
by $e_{\nu\nu}^{z}(\bdq)$, respectively,
we can express $L_{11}^{\prime}$ and $L_{22}^{\prime}$
in the limit $\tau\rightarrow \infty$ as follows:
\begin{align}
  L_{11}^{\prime}=&\frac{\pi}{N^{2}}\sum_{\bdq,\bdq^{\prime}}
  \sum_{\nu,\nu^{\prime},\nu^{\prime\prime}=\alpha,\beta}
  v_{\nu\nu}^{z}(\bdq)v_{\nu^{\prime}\nu^{\prime}}^{z}(\bdq^{\prime})\tau^{2}
  \frac{\partial n[\epsilon_{\nu}(\bdq)]}{\partial \epsilon_{\nu}(\bdq)}
  S\sin^{2}2\phi \sum_{p=1,2,3}
  F^{(p)}_{\nu\nu^{\prime}\nu^{\prime\prime}}(\bdq,\bdq^{\prime}),\label{eq:L11'-last}\\
  L_{22}^{\prime}=&\frac{\pi}{N^{2}}\sum_{\bdq,\bdq^{\prime}}
  \sum_{\nu,\nu^{\prime},\nu^{\prime\prime}=\alpha,\beta}
  e_{\nu\nu}^{z}(\bdq)e_{\nu^{\prime}\nu^{\prime}}^{z}(\bdq^{\prime})\tau^{2}
  \frac{\partial n[\epsilon_{\nu}(\bdq)]}{\partial \epsilon_{\nu}(\bdq)}
  S\sin^{2}2\phi \sum_{p=1,2,3}
  F^{(p)}_{\nu\nu^{\prime}\nu^{\prime\prime}}(\bdq,\bdq^{\prime}).\label{eq:L22'-last}
\end{align}
Equations (\ref{eq:L12'-last}), (\ref{eq:L11'-last}), and (\ref{eq:L22'-last})
yield Eq. (\ref{eq:L'}). 
\end{widetext}

\section{Additional numerical results of $S_{\textrm{m}}$, $\sigma_{\textrm{m}}$, and $\kappa_{\textrm{m}}$}

\begin{figure*}
  \includegraphics[width=180mm]{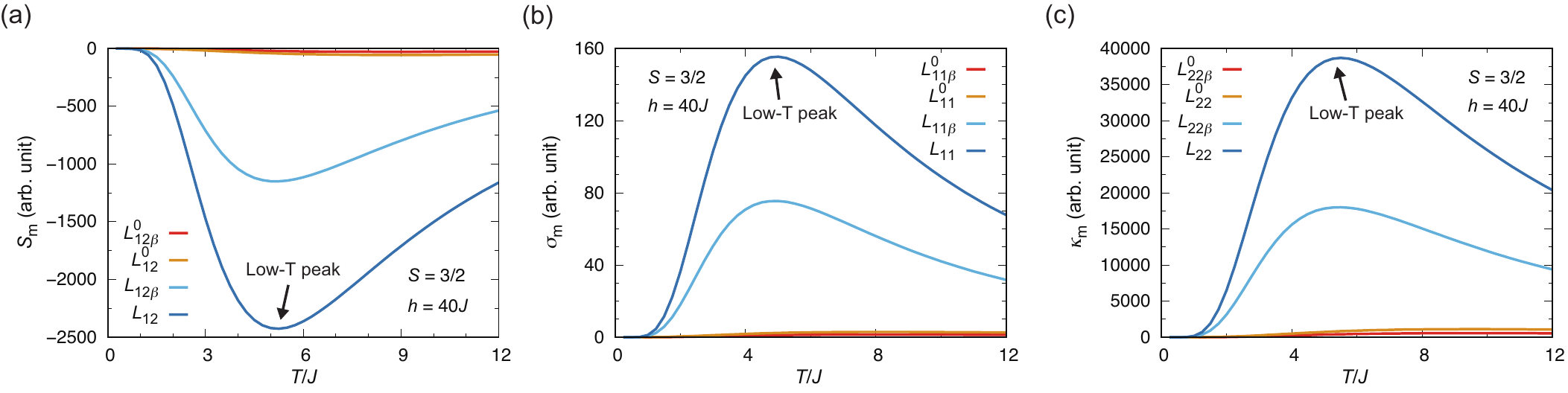}
  \caption{\label{fig5}
    The temperature dependences of (a) $S_{\textrm{m}}$,
    (b) $\sigma_{\textrm{m}}$, and (c) $\kappa_{\textrm{m}}$
    obtained in the numerical calculations for $S=\frac{3}{2}$
    with $\frac{N}{2}=20^{3}$ and $J=1$
    at $h=40J$.
    The red, yellow, light blue, and blue curves
    represent the $T/J$ dependences of $S_{\textrm{m}}=L_{12\beta}^{0}$,
    $\sigma_{\textrm{m}}=L_{11\beta}^{0}$, and $\kappa_{\textrm{m}}=L_{22\beta}^{0}$,
    those of $S_{\textrm{m}}=L_{12}^{0}$,
    $\sigma_{\textrm{m}}=L_{11}^{0}$, and $\kappa_{\textrm{m}}=L_{22}^{0}$,
    those of $S_{\textrm{m}}=L_{12\beta}$,
    $\sigma_{\textrm{m}}=L_{11\beta}$, and $\kappa_{\textrm{m}}=L_{22\beta}$,
    and
    those of $S_{\textrm{m}}=L_{12}$,
    $\sigma_{\textrm{m}}=L_{11}$, and $\kappa_{\textrm{m}}=L_{22}$,
    respectively.
    $L_{\mu\eta\beta}^{0}$ is part of the noninteracting term,
    the contribution from the lower-branch magnons (i.e., the $\beta$-band magnons); 
    $L_{\mu\eta}^{0}$ and $L_{\mu\eta}^{\prime}(=L_{\mu\eta}-L_{\mu\eta}^{0})$ are
    the noninteracting and drag terms, respectively.
    $L_{\mu\eta\beta}=L_{\mu\eta}^{0}+L_{\mu\eta\beta}^{\prime}$,
    where $L_{\mu\eta\beta}^{\prime}$ is part of the drag term,
    the contribution from
    the term for $(\nu,\nu^{\prime},\nu^{\prime\prime})=(\beta,\beta,\beta)$
    in Eq. (\ref{eq:L'}).
  }
\end{figure*}

We present the additional results of the numerically evaluated
$S_{\textrm{m}}$, $\sigma_{\textrm{m}}$, and $\kappa_{\textrm{m}}$
for $S=\frac{3}{2}$ with $\frac{N}{2}=20^{3}$ and $J=1$.
(In the case of $S=\frac{3}{2}$,
the magnon picture for the canted antiferromagnet is valid
in the range of $0<h<48J$.)
Since the transition temperature for $S=\frac{3}{2}$ becomes $T_{\textrm{c}}=20J$,
we choose the temperature range to be $0<T\leq 12J(=0.6T_{\textrm{c}})$. 
Figures \ref{fig5}(a){--}\ref{fig5}(c) show
the temperature dependences of $S_{\textrm{m}}$, $\sigma_{\textrm{m}}$, and $\kappa_{\textrm{m}}$
for $S=\frac{3}{2}$ at $h=40J$.
For $S=\frac{3}{2}$, 
the low-temperature peaks are observed at the $h$ lower than $65J$.
Then, the ratios $L_{12}/L_{12}^{0}$, $L_{11}/L_{11}^{0}$, and $L_{22}/L_{22}^{0}$
at $T=5J(=0.25T_{\textrm{c}})$ reach about $60$, $66$, and $52$, respectively.
The larger enhancement for $S=\frac{3}{2}$ than that for $S=\frac{5}{2}$
comes from the property that 
the smaller the $S$ is,
the more considerable 
the effects of magnon-magnon interactions become.
This general property is due to the difference between
the $S$ dependences of $H_{0}$ and $H_{\textrm{int}}$.

\end{document}